\documentclass[12pt,preprint]{aastex61}
\usepackage{amsmath}
\usepackage{natbib}
\usepackage{hyperref}
\usepackage{xspace}

\usepackage{algpseudocode}

\DeclareFontFamily{U}{wncy}{}
\DeclareFontShape{U}{wncy}{m}{n}{<->wncyr10}{}
\DeclareSymbolFont{mcy}{U}{wncy}{m}{n}
\DeclareMathSymbol{\Sha}{\mathord}{mcy}{"58} 

\DeclareFontFamily{U}{wncy}{}
\DeclareFontShape{U}{wncy}{m}{n}{<->wncyr10}{}
\DeclareSymbolFont{mcy}{U}{wncy}{m}{n}
\DeclareMathSymbol{\Shah}{\mathord}{mcy}{"58} 

\DeclareMathOperator{\corr}{corr}

\DeclareMathOperator{\acf}{ACF}
\DeclareMathOperator{\pacf}{PACF}

\shorttitle{AutoRegressive Planet Search}
\shortauthors{Caceres et al.}

\begin{document}

\title{AutoRegressive Planet Search: Methodology}
\author{Gabriel A. Caceres}
\affiliation{Department of Astronomy \& Astrophysics, Pennsylvania State University, University Park, PA 16802}
\affiliation{SparkBeyond, 270 Madison Ave., Suite 702, New York NY 10016}

\author{Eric D. Feigelson}
\altaffiliation{Center for Exoplanets and Habitable Worlds, Pennsylvania State University, University Park PA 16802}
\affiliation{Department of Astronomy \& Astrophysics, Pennsylvania State University, University Park, PA 16802}
\affiliation{Center for Astrostatistics, Pennsylvania State University, University Park PA 16802}
\email{edf@astro.psu.edu}

\author{G. Jogesh Babu}
\affiliation{Department of Statistics, Pennsylvania State University,  University Park, PA 16802}
\affiliation{Center for Astrostatistics, Pennsylvania State University, University Park PA 16802}

\author{Natalia Bahamonde and Alejandra Christen}
\affiliation{Instituto de Estad\'{i}stica, Pontificia Universidad Cat\'{o}lica de Valpara\'{i}so, Valpara\'{i}so, Chile}

\author{Karine Bertin and Cristian Meza}
\affiliation{Centre for Research and Modeling of Random Phenomena, CIMFAV, Universidad de Valpara\'{i}so, Valpara\'{i}so, Chile}

\author{Michel Cur\'{e}}
\affiliation{Instituto de F\'{i}sica y Astronom\'{i}a, Universidad de Valpara\'{i}so, Valpara\'{i}so, Chile}

\begin{abstract}
The detection of periodic signals from transiting exoplanets is often
impeded by extraneous aperiodic photometric variability, either
intrinsic to the star or arising from the measurement
process. Frequently, these variations are autocorrelated wherein
later flux values are correlated with previous ones. In this work, we
present the methodology of the Autoregessive Planet Search (ARPS)
project which uses Autoregressive Integrated Moving Average (ARIMA)
and related statistical models that treat a wide variety of stochastic processes, as
well as nonstationarity, to improve detection of new planetary
transits. Providing a time series is evenly spaced or can be placed on
an evenly spaced grid with missing values, these low-dimensional parametric models
can prove very effective. We introduce a planet-search algorithm to
detect periodic transits in the residuals after the application of
ARIMA models. Our matched-filter algorithm, the Transit Comb Filter
(TCF), is closely related to the traditional Box-fitting Least Squares
and provides an analogous periodogram. Finally, if a previously
identified or simulated sample of planets is available, selected scalar
features from different stages of the analysis -- the
original light curves, ARIMA fits, TCF periodograms, and folded light
curves -- can be collectively used with a multivariate classifier to
identify promising candidates while efficiently rejecting false alarms. We use
Random Forests for this task, in conjunction with Receiver Operating
Characteristic (ROC) curves, to define discovery criteria for new,
high fidelity planetary candidates. The ARPS methodology can be
applied to both evenly spaced satellite light curves and densely
cadenced ground-based photometric surveys.
\end{abstract}


\accepted{for the Astronomical Journal, 2019 May 14}

\section{Introduction} \label{sec:Intro}

Searching for transits in stellar light curves has been a very
successful approach to discovering exoplanets, yielding several
thousand candidates to date. NASA's \textit{Kepler} Mission has
identified a significant fraction of known exoplanets with densely
cadenced, high-precision photometry over nearly 4
years~\citep{Borucki2010}.  One of the main challenges facing
\textit{Kepler} and other spaced-based planet discovery surveys is the
noise and variability of stars with amplitudes comparable to or exceeding
the depth of planetary transits~\citep{Gilliland2011}. Aperiodic `red 
noise' is particularly prevalent and 
challenging to treat \citep{Pont2006, Carter2009, Cubillos2017}.  Ground-based
surveys suffer a similar problem but with extraneous variability
arising predominantly from instrumental and atmospheric conditions\footnote{
  For simplicity, in the
  remainder of this paper we will refer to these sources of extraneous
  variability as 'stellar', recognizing that other sources may be
  present.}.
Further discoveries will greatly benefit from statistical methods capable of
recovering fainter planetary signals, such as Earth analogues, in
systems with high variability.

The problem of identifying transiting planets from photometric time
series can be viewed in three stages: (1) a time-domain regression, transform, or
interpolation procedure to identify and remove stellar variability;
(2) a frequency-domain procedure to construct periodograms in order to
find orbital behaviors in the residual time series; and (3) a way to
discriminate planetary transit candidates from statistical false alarms and
astronomical false positives, such as a
decision tree based on various signal statistics.

For the first stage, most researchers use nonparametric approaches
including wavelet analysis~\citep{Jenkins2002,Carter2009}, 
Fourier filtering~\citep{Carpano2003, Huang2013}, local
linear modeling~\citep{Roberts2013},  Gaussian Processes
regression~\citep{Gibson2014, Aigrain2016, Luger2016}, 
Independent Components Analysis~\citep{Waldmann2013}, and
Singular Spectrum Analysis~\citep{Boufleur2018}. The Kepler Team 
provides {\it PyKE}, a detrending procedure based on iterative local polynomial fitting 
\citep{Vinicius17}.  Here, \textit{nonparametric} refers to models where no functional
relationship is globally applied to the time series, although
semi-parametric regressions may be used locally~\citep{Ruppert2003,Takezawa2005}.

For the second stage, the most widely used tool is the Box-fitting
Least Squares (BLS) algorithm of \citet{Kovacs2002} that acts as a
matched filter for box-shaped planetary transits. Other procedures for
periodicity searches include periodograms from phase dispersion
minimization~\citep{Plavchan2008}, Fourier~\citep{Sanchis-Ojeda2014}
and Lomb-Scargle~\citep[LS; e.g.,][]{Hartman2008} transforms.  See 
\citet{Graham13b} for a comparison of period-finding methods. 

The third stage to distinguish candidate transits from other related
periodic behaviors is typically based on the presence of strong peaks
in the periodogram and a variety of additional quantitative and
qualitative considerations falling under the rubric of `vetting'.
Automated vetting of planetary transits can utilize procedures such as 
decision trees and Random Forests \citep{McCauliff2015, Mislis2016, Armstrong2018}; 
these are well-established techniques from modern machine learning 
methodology~\citep{Breiman2001}\footnote{We do not consider here procedures such 
as the trend filtering \citep{Kovacs2005}, {\it Sysrem} \citep{Tamuz2005}
and {\it SARS} \citep{Ofir2010} algorithms that treat collective 
variations in ensembles of nearby stars.  In this study, each light curve is 
assumed to be independent of other light curves. The AutoRegressive 
Planet Search procedure is best run after collective effects from instrumental 
or atmospheric conditions are removed as much as possible. \label{footnote2}}.

While stellar variability can arise from eclipses, pulsations, and
other phenomena, it is most commonly due to magnetic activity
including photospheric starspots, chromospheric plages, and
reconnection flares~\citep{Schrijver2000}.  A particular property of solar and
stellar activity is autoregressive behavior, wherein future
photometric values depend on current and past values.  For example,
solar flare occurrences are often modeled as `avalanche' processes
that produce a $1/f$-type behavior responsible for power law
statistical distributions in solar activity properties~\citep{Lu1991,
  Aschwanden2016}. The frequency distribution of solar and stellar
flares is a power law over 10 orders of magnitude in energy.  We will
see below that $1/f$-type `long memory' processes can be treated by
certain autoregressive models~\citep{Palma2007}.

The autocorrelated characteristics of stellar photometric `noise'
leads naturally to the idea that stochastic autoregressive statistical
models could be applied to reduce their influence and help reveal
faint periodic planetary transits.  The simple case of an
autoregressive moving average is labeled an `ARMA' model.  ARMA-type
statistical models were popularized by \citet[][the latest edition
is Box et al.\ 2015]{Box1970} and are now very
widely used in signal processing, voice recognition, and econometrics,
among many other applications.  These are \emph{parametric} models
where the coefficients quantify the dependency of current values on
past ones assuming stationarity (where the behavior is unchanged
throughout the time series). While simple ARMA models treat only
`short-memory' autocorrelated processes and white noise in stationary
time series, more elaborate models allow for non-stationarity and
`long-memory' $1/f$-type processes.

As ARMA-type models are low-dimensional global parametric regression
models---rather than nonparametric or high-dimensional semi-parametric
models ---powerful likelihood-based statistical regression procedures
can be utilized.  The common procedure is to compute maximum
likelihood best-fit models, and then choose the most parsimonious one
consistent with the data using penalized likelihood measures such as
the Akaike Information Criterion \citep{Hamilton1994,Chatfield2004}. Quantitative
automated measures can be used, leaving these procedures with no
arbitrary free parameters or subjective choices.  In contrast, nonparametric 
procedures are subject to choices:  the smoothing
kernel function and bandwidth in Gaussian Processes regression;  
the basis function, denoising threshold, and band selection in wavelet analysis; and so forth. 
Well-established statistical goodness-of-fit tests are available to test whether 
the best model does indeed fit the data well.  

Here we develop a three-stage AutoRegressive Planet Search (ARPS) procedure
in detail.  We start with maximum likelihood fits of \textit{integrated} (ARIMA)
and \textit{fractionally integrated} (ARFIMA) extensions to the ARMA model,
reducing unwanted photometric variability.  This is described in \S\ref{sec:ar-models}.  
Simple ARMA models have been 
discussed elsewhere for limited aspects of photometric planet searches 
\citep{Carter2009, Wang2016} and for filling gaps in irregular light curves
\citep{Fahlman1982, Pascual2015}.  The richer ARIMA and ARFIMA families 
are largely absent from astronomical studies though have considerable
potential \citep{Feigelson2018}. 

However, the temporal
nature of the transits are transformed by the modeling from a periodic
box-like transit shape to a periodic double-spike-like shape.  This
required development of a customized matched filtering algorithm,
called here the `Transit Comb Filter' (TCF), to construct
periodograms as presented in \S\ref{sec:tcf}.  

In the final stage, `features' from the light curve, ARIMA-type fits, 
TCF periodograms, and other aspects of the analysis are fed into a
machine learning classifier to recover known, and discover new, 
candidate planet transit systems (\S\ref{sec:class}).  The Random
Forests extension of multivariate decision tree classification is an
effective method for this stage.  Receiver Operating Characteristic 
(ROC) curves help select thresholds for discrimination of promising
planet candidates from false alarm and false positive signals. 

ARPS methodology thus 
differs from common planet finding procedures
in the following ways: we use principally ARIMA (instead of moving
medians, wavelets, or Gaussian Processes regression) to model the
star; the TCF periodogram (instead of BLS or LS periodogram) to extract
the periodic signals from transits; and \citep[similar to][]{McCauliff2015}
multivariate classification using Random Forests (instead of a univariate 
measure like periodogram power).   Various methodological issues
are discussed in \S\ref{sec:disc} with conclusions in \S\ref{sec:concl}. 

The present study describes the mathematics and analysis procedure
underlying the ARPS project, giving examples from observed
\textit{Kepler} light curves.  A first companion
paper~\citep{Caceres2019b} will apply the method to the full sample of
$\sim 200,000$ stellar light curves from NASA's \textit{Kepler}  mission.  Using the final data release
DR-25, these light curves span $\sim$70,000 evenly spaced time stamps
with $\sim 15-20\%$ of the time stamps have missing data due to instrumental causes.
A second companion paper \citep{Stuhr2019} investigates through simulation
the applicability of ARPS to irregularly spaced ground-based transit surveys.
Further studies are in progress applying the methods to data from the space-based
\textit{TESS} mission and ground-based HATSouth survey.  

Throughout this paper, we will return to the two \textit{Kepler} stellar light curves
shown in Figure~\ref{fig:lc-ex} to illustrate the various stages of
analysis on realistic astrophysical light curves.  The first star has
only low-level variability, hardly discernible above the white noise
of the instrument.  The second star has high amplitude variability far
larger than the noise.  The notation `IQR' stands for interquartile
range, a robust nonparametric measure of spread analogous to the
standard deviation for Gaussian distributions.

\section{Autoregressive Modeling}
\label{sec:ar-models}

\subsection{Overview}

When a dynamic system varies over time, it is common that its
current state depend on its past behavior, in which case the process is said to
be autocorrelated. The evolution of such a system need not follow a
deterministic path. As might seem natural, time series of many
physical processes display stochastic autocorrelated properties. This
is often the case with variations in stellar photometry, and thus the
key motivation for our approach since a great variety of parametric
models have been developed to model these aperiodic, stochastic,
temporal behaviors. These models weight the influence of past measured
values, not (only) current values as in more common regression
situations. 

The textbooks of~\citet{Box2015},~\citet{Chatfield2004}, 
and~\citet{Shumway2006}, are useful general references and provide 
further details on the topics introduced throughout this section.  
Mathematically advanced treatments of ARIMA and ARFIMA models 
appear in volumes by~\citet{Hamilton1994}, \citet{Palma2007} and~\citet{Beran2013}.
Additionally,~\citet{Scargle1981} and~\citet{Koen1993} give 
valuable reviews of these topics oriented towards astronomers.

The utility and power of statistical models and theorems rests on the
validity of the underlying assumptions for a given problem.  Many
statistical inferential procedures, such as ordinary least squares,
assume independent and identically distributed (i.i.d.) random
variables.  But, where autocorrelation is present, independence is
absent.  Care must be taken with conclusions derived from analyzing
dependent data. It has long been known that correlated errors have an
impact on statistical modeling, and the effects have been extensively
studied~\citep[see][for a review of early work]{Anderson1954}. For
example,~\citet{Cochrane1949} showed that when the errors are
serially correlated, the least\nobreakdash-squares estimates are no
longer guaranteed to have minimum-variance---though they may still
remain unbiased. Furthermore, the usual estimates of variance (and
thus standard errors) do not apply, and the use of t and F
distributions for confidence calculations are no longer
valid~\citep{Durbin1950}. \citet{Romano1996} also discuss
problems that arise with inference of autocorrelation and ARMA
coefficients when the underlying assumptions are not strictly
satisfied.

We will avoid spending significant effort addressing these mathematical issues by
approaching the models from a more phenomenological point of view. The goal here
is to reduce the presence of autocorrelation in order to discover
new candidate transiting planets, but there is no need or expectation to calculate the ``true''
ARMA coefficients underlying the physical phenomena, nor even make the
assumption it is the intrinsically ``correct'' model.

The principal focus of this work are Autoregressive (Fractionally)
Integrated Moving Average models, known as ARIMA and ARFIMA. In broad
terms, AR(F)IMA models represent a regression of the data as a
function of its past state.  The AR and MA components reflect the
dependence of current values on recently past values, fractional
integration reflects long timescale dependencies such as $1/f^\alpha$
noise, while integer values reflect drifts in the mean.  In the
parlance of time series analysis, these three components treat
short-memory processes, long-memory processes, and nonstationarity,
respectively.  These approaches have seen widespread, successful use
in other fields ranging from engineering to econometrics.  The
iterative approach to analyze and apply these models was popularized
by the seminal work of Box~\&~Jenkins.

Following standard presentations in time series analysis, we assume
that the data are acquired at evenly spaced intervals, although
missing data at some time slots is permitted.  Discrete measurements of the 
temporal process $X(t)$ produce a sequence of observations $x_t$ where 
$t=1, 2,\ldots, n$.  The \textit{Kepler} long-cadence photometric data are
acquired in evenly space 29.4 minute intervals, but many other
astronomical datasets are unevenly spaced.  

Though not as pervasive as in other fields,  simple ARMA-type models are increasingly used 
in time domain astronomy, currently around 25 studies annually \citep{Feigelson2018}.   
The richer classes of ARIMA and ARFIMA models that treat non-stationarity and long-memory
`red' noise as well as short-memory processes only rarely appear in astronomical studies.
ARFIMA is used by \citet{Stanislavsky2009} to characterized solar flaring in the X-ray band. 
For irregularly spaced data, the model can be reconfigured as a continuous-time process.  
Simple CAR and CARMA models are used for analysis of quasar light curves \citep{Kelly2014}, 
but broader CARFIMA models \citep{Tak2017} have yet to be applied.   \citet{Eyheramendy2018}
develop an important generalization, nicknamed IAR for the Irregular AutoRegressive model, that
treats non-Gaussian errors. 

We note that autoregressive modeling is most effectively used after systematic variations due to instrumental or atmospheric conditions are reduced.  Widely used algorithms for this problem include the Transit Filter Algorithm \citep{Kovacs2005}, the SysRem algorithm \citep{Tamuz2005}, and Presearch-Data Conditioning \citep{Stumpe2012}.  But ARIMA-type techniques can be used even when systematic effects are not fully removed.  The mathematics makes no distinction between autocorrelated variations intrinsic to the star from those arising from the observational process.

\subsection{Time Series Diagnostics}
\label{sec:acf}

\begin{figure}[t]
  \begin{center}
    \includegraphics[width=\textwidth]{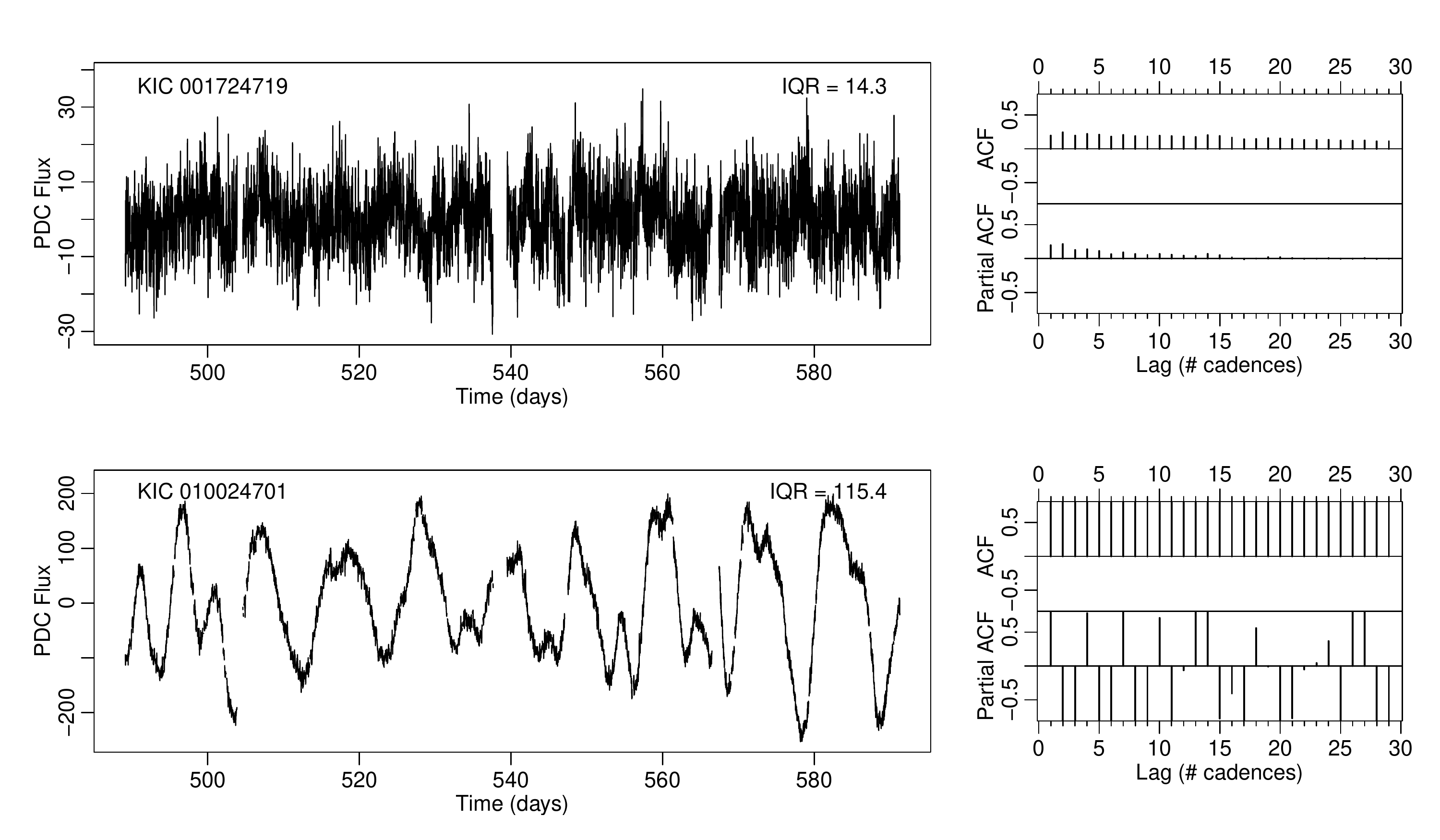}
  % \plotone{figures/lc-ex}
  \caption{Example time series and corresponding (P)ACF for KIC 001724719 (Kepler-1569 with P=5.79 day transit period) and KIC 010024701 (KOI K02002.01 with P=14.4 day).  We show here only a few percent of the full Kepler light curve that extends for $\sim 1500$ days. The ordinate shows the median-subtracted stellar flux in electrons per second after application of the Kepler Team Pre-Search Data Conditioning Pipeline module that removes most instrumental effects.  IQR gives  the Interquartile Range of PDC Flux for the full light curve.  The autocorrelation function plots are truncated above 0.5.   \label{fig:lc-ex}}
  \end{center}
\end{figure}

The need for autoregressive modeling can be determined by evaluating
the presence of correlated noise in the time series under study.  The
Autocorrelation Function (ACF) is a fundamental nonparametric measure
of autocorrelation in stationary time series.  It calculates the
degree of correlation between a series and time-lagged values of
itself over the entire time series.  Like the Fourier and wavelet
transforms, all information in a time series is maintained in the ACF
if an unlimited number of coefficients are kept.  The advantage
of the ACF is that it concentrates short-memory autocorrelation into a
few coefficients, even if it is distributed weakly throughout the time
series. The Partial Autocorrelation Function (PACF) is a variant of
the ACF that measures the lagged correlation while controlling for the
effects contributed by intermediate lags.

The cross-correlation between two time series is defined by
\begin{equation}
  \label{eq:corr}
  \corr(x,y) = \frac{E[(x - \mu_x)(y - \mu_y)]}{\sigma_x\sigma_y}
\end{equation}
where $x$ and $y$ are two time series with means $\mu_i$ and standard
deviations $\sigma_i$, and $E$ is the expected value. The
ACF at lag $k$ corresponds to the correlation between the random variables $x_t$ and $x_{t-k}$, and can thus be written as
\begin{equation}
  \label{eq:acf}
  \acf(k) = \corr(x_t,x_{t-k}) = \frac{E[(x_t - \mu_x)(x_{t-k} - \mu_x)]}{\sigma_x^2}
\end{equation}
Note that at lag 0 the ACF always equals 1 since the numerator simply corresponds to the constant variance of the series.

While the ACF gives a measure of the memory of the process, the PACF seeks to identify the direct individual contribution from the $k$-th lag, removing the effects of the other intermediate lags. 
The PACF for a stationary process is given by
\begin{eqnarray}
  \label{eq:pacf}
  \pacf(1) &=& \corr(x_1,x_0) = \acf(1) \\
  \pacf(k) &=& \corr(x_t, x_{t-k} | x_{t-1}, \dotsc, x_{t-(k-1)}) \quad k \geq 2 \nonumber
\end{eqnarray}
where these coefficients are estimated by fitting autoregressive models of successively higher orders.  Examples of ACFs and PACFs appear in Figures~\ref{fig:lc-ex}-\ref{fig:lc-arma}.

In addition to the ACF and PACF, hypothesis tests exist to evaluate whether a sequence of data has correlated noise.  The Durbin\nobreakdash-Watson test~\citep{Durbin1950, Durbin1951} measures serial (lag=1) autocorrelation, and the  Ljung\nobreakdash-Box test~\citep{LjungBox1978} is a portmanteau test for all lags;  they are commonly applied to the residuals of a regression model. Related tests include: the Anderson-Darling and Jarque-Bera tests for normality; the Breusch-Pagen and White tests for heteroscedasticity; and the Augmented Dickey-Fuller and Kwiatkowski-Phillips-Schmidt-Shin tests for stationarity.  These are described in textbooks for econometrics~\citep{Enders2014, Greene2017, Hyndman2014}.   We also make use of the Breusch-Godfrey test~\citep{Breusch1978,Godfrey1978} which generalizes the Durbin-Watson test for lags $k>1$.

\subsection{(Non)Stationarity and the Differencing Operator}
\label{sec:diff}

\begin{figure}[t]
  \begin{center}
    \includegraphics[width=\textwidth]{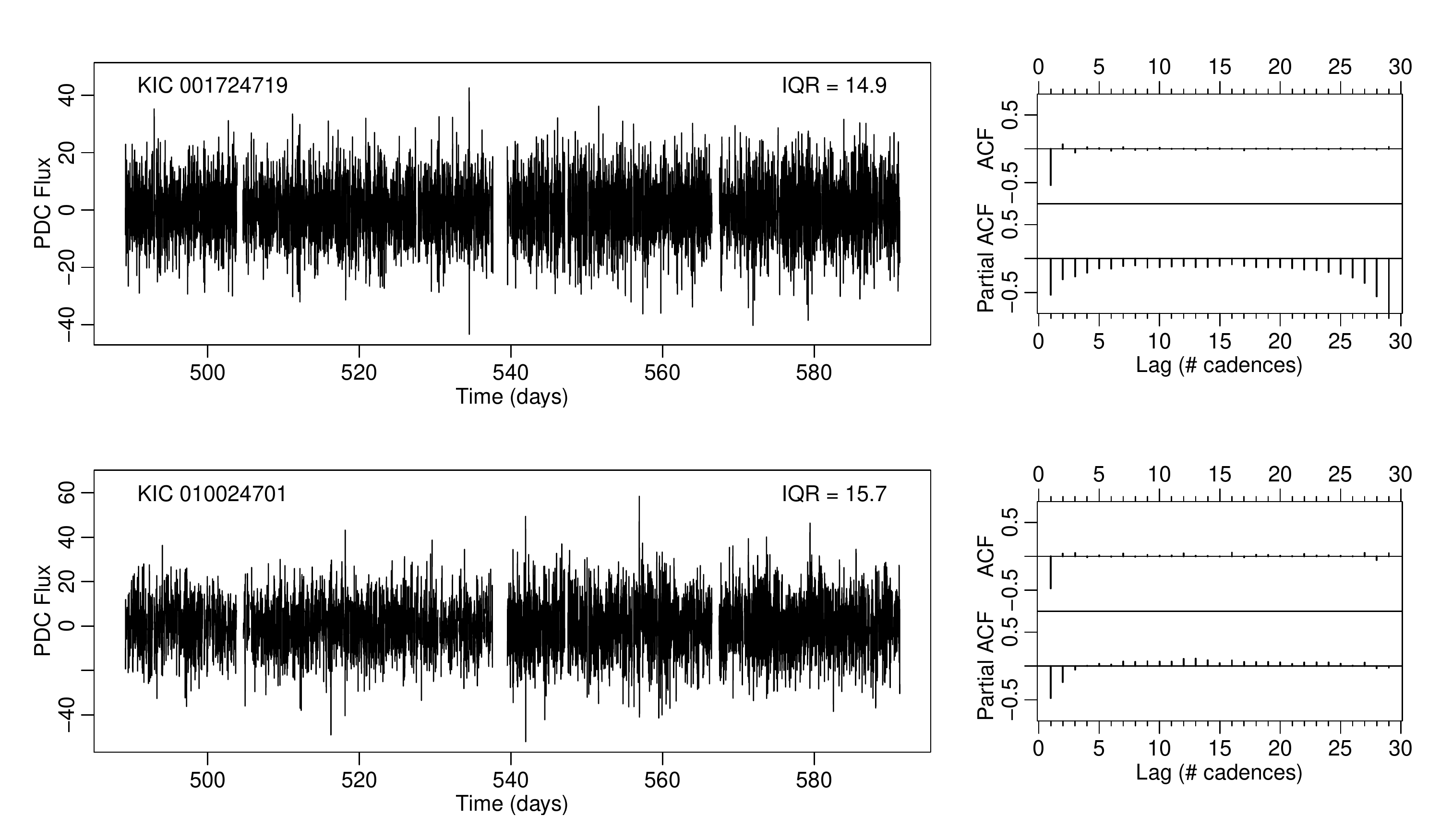}
  \caption{Time series and corresponding (P)ACF after differencing operation is applied to the lighturves shown in Figure~\ref{fig:lc-ex}.  Note the strong, but not complete, reduction in noise and autocorrelation compared to Figure~\ref{fig:lc-ex}. 
  \label{fig:lc-diff}}
  \end{center}
\end{figure}

A time series is stationary when its properties do not change over
time, so that its global characteristics (such as mean, variance, and ACF) are identical irrespective of when it is observed. Specific local values will, of course, vary
between different segments of time. More formally, strongly stationary
processes have a joint probability distribution that is
independent of time. ARMA models require wide-sense (or weak) stationarity, where the first two moments---the mean and autocovariance---remain approximately constant over time and the correlation structure depends only on the lag of the observations.

Stationarity is violated when a trend in the mean is present.  But nonstationarity can also be present in stochastic processes when the system does not revert to the mean when subject to random shocks.  The physicist's `random walk' is nonstationary in this sense; mathematically, the time series is said to have a unit root.  Trends can be reduced by (local) regression techniques, and unit root nonstationarity can be reduced by applying the differencing operator (below).  Strictly periodic variations, however, are a type of nonstationarity that is not effectively treated in this fashion and should be modeled using frequency domain techniques.  

Aperiodic cyclic behaviors, often called quasi-periodicities,  
can still arise from a stationary stochastic process even when there is no physical origin for periodicity such as stellar rotation or planetary orbit.  Indeed, ARMA-type processes often produce quasi-periodicities that change or dissipate on long timescales;  see \citet{Vaughan2016} for an astronomical perspective.  Long-memory $1/f^\alpha$ autocorrelated processes can
be stationary or nonstationary depending on the value of $\alpha$. Time
series where the mean values drift or non-recurring
outbursts are also nonstationary. More generally, any deterministic
function $X=f(t)$ can cause nonstationarity. 

Astronomical time series are often nonstationary. A Mira variable star exhibits 
long-term nonstationary trends in brightness.    A solar-type star can exhibit quasi-periodic 
variations due to rotationally modulated starspots.  The X-ray emission
from the accreting stellar black hole binary GRS 1915+105 is a famous
example where the source shifts between more than a dozen different modes of
variations~\citep{Belloni2000}. Astronomers commonly trace these
behaviors with a local regression procedure, such as a moving average  or
Gaussian Processes regression, or with wavelet analysis.  But a very simple,
and often effective, method for reducing many forms of nonstationarity is through differencing. 
For the backshift operator $B$ defined as 

\begin{equation} 
\label{eq:diff0}
Bx_t = x_{t-1}
\end{equation}

and the resulting differenced series is given by 

\begin{equation}
\label{eq:diff1}
  x^{\prime}_t=x_t-Bx_t=x_t-x_{t-1}
\end{equation}

This is a transform to address the presence of nonstationarity, much
in the same manner as \texttt{log} transforms are used to deal with
wide ranges and heteroskedasticity. 
One can also view the differencing operation as a
high-pass filter which removes the long-scale variations while leaving
short-term fluctuations.  The operator takes the point-to-point
difference of the original data (\emph{i.e.} $x_t - x_{t-1}$), to
create a new stationary time series. This new series of the
\emph{changes} in the data can then be modeled with the autoregressive
methods further described in \S\ref{sec:arima}. 

Formally,
differencing a series is meant to only address certain types of
nonstationarity, like a stochastic trend component.
In practice a single application of the
differencing operator to a nonstationary time series is often
sufficient to render it (approximately) stationary irrespective of the 
exact nature of the nonstationarity.  Differencing can increase
noise levels if the original time series is Gaussian white noise, but 
usually decreases noise when autocorrelation is present.

Figure~\ref{fig:lc-diff} shows the effects of differencing on the two Kepler light curves
shown in Figure~\ref{fig:lc-ex}.

\subsection{Modeling Autocorrelated Data}
\label{sec:arima}

\begin{figure}[t]
  \begin{center}
    \includegraphics[width=\textwidth]{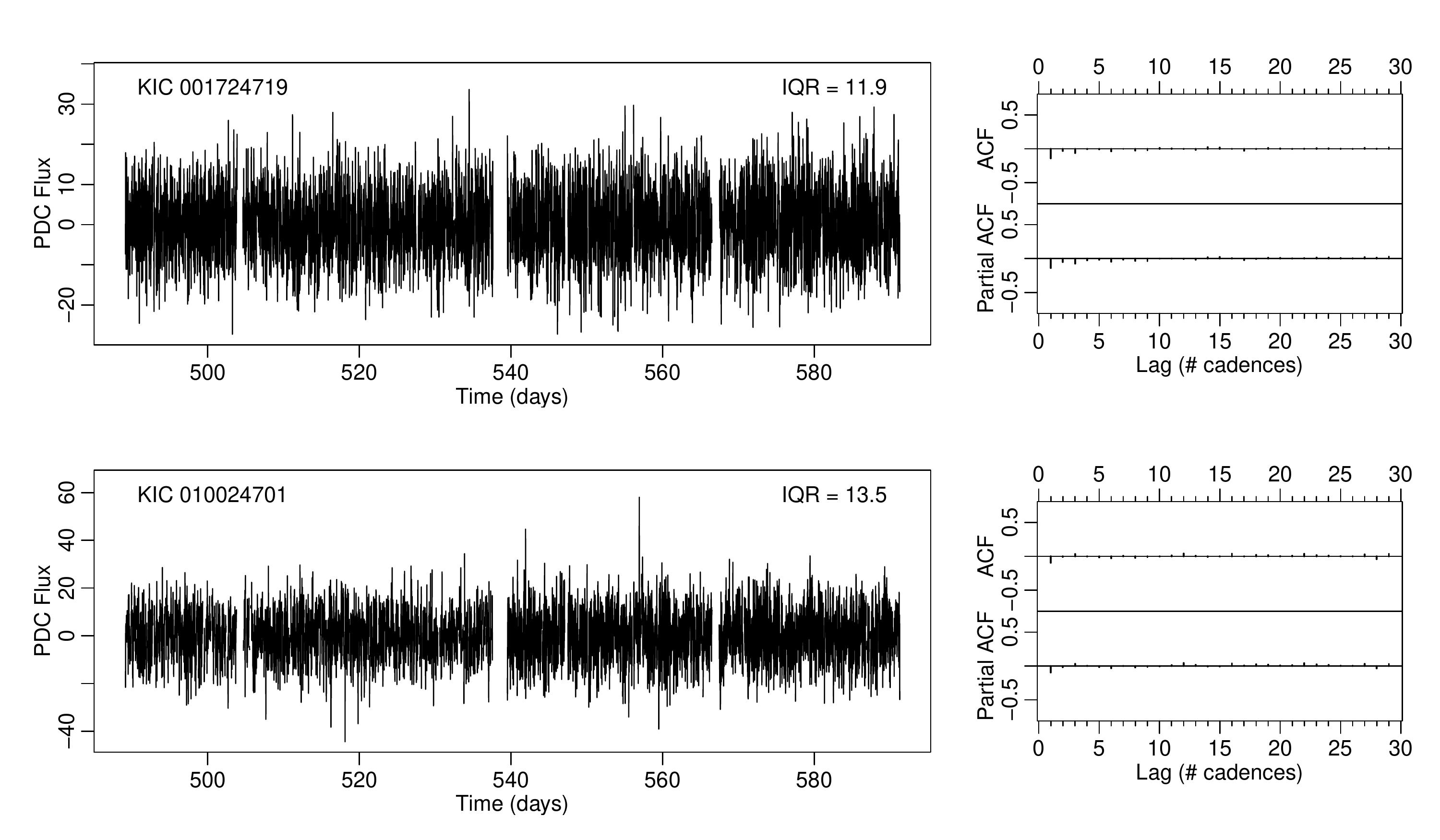}
  \caption{Residual time series and corresponding (P)ACF after autoregressive modeling is applied to the data shown in Figure~\ref{fig:lc-diff}. The residuals show no significant autocorrelation and are close to Gaussian white noise.}
  \end{center}
  \label{fig:lc-arma}
\end{figure}

The ACF (\S2.1) and differencing operators (\S2.2) are $nonparametric$
transforms of stationary and nonstationary time series, respectively.
The ARPS analysis is centered on $parametric$ modeling of
autocorrelated time series, treating both stationary and nonstationary
cases.  

We consider the large class of ARMA-type models where ARMA is
an acronym for `autoregressive moving average'.  The class is very
broad with important variants such as VARMA (`vector' ARMA for
multivariate time series with lags), CARMA (`continuous' time ARMA for
interpolating irregularly spaced observations), and SARMA (`seasonal'
ARMA with a strictly periodic components). There are also models like ARCH and 
GARCH (`generalized autoregressive conditional heteroskedasticity') to
treat data with non-constant, autocorrelated variance.  The 2003 Nobel Prize 
in Economic Sciences was awarded to Robert Engle for development of the 
ARCH model that treats volatility where the noise values are also parametrized 
as an autoregressive process.  In the ARPS
analysis, we concentrate on two ARMA variants: Autoregressive
Integrated Moving Average (ARIMA) models which can deal with nonstationarity 
and, and ARFIMA models which also treat long-memory processes by allowing
\textit{fractional} integration. 

These models serve to phenomenologically characterize the
behavior of an autocorrelated stochastic process, and in some
contexts, predict its evolution.  In many situations, apparently
complicated and stochastic time series are actualizations of
relatively simple models that depend on only a few parameters
describing the autocorrelated behavior. Of key importance is that these
methods are designed to model stochastic processes where the random
component of the series has a direct effect on its evolution due to the correlated structure.

\subsubsection{Autoregressive (AR) Process:\label{sec:ar}}

If the value of a variable at a given point in time is influenced by
its past values, the model is said to be autoregressive; sequential
observations in this scenario are not independent. To account for this
effect, a stationary time series can be modeled as a linear combination of its
lagged values plus an additional random noise term. A pure \emph{AR} process can be seen as simple regression problem,
with the current state represented as a linear combination of past values.
An \emph{AR(p)} process is modeled 
by
\begin{equation}
  \label{eq:ar}
  x_t = \phi_1 x_{t-1} + \phi_2 x_{t-2} + \ldots + \phi_p x_{t-p} + \epsilon_t 
\end{equation}
where $\epsilon_t$ is a normally (Gaussian) distributed random error 
with zero mean and unknown variance, $\epsilon_t = N(0,\sigma^2)$, 
$p$ is the order of the process
(\emph{i.e.} the number of lags in the model), and $\phi_i$ are the
corresponding coefficients for each lag up to order $p$. The values of $\phi_i$ can thus be
calculated, for example, via least squares or maximum likelihood estimation.
Additionally, the sum of the parameters are constrained
to be less than one for stationary processes.  
An artificial case where all $\epsilon_t$ values are zero  can mimic a strictly periodic time series.

AR processes have a close link to the ACF and PACF discussed in \S\ref{sec:acf}, with
each AR model having a characteristic structure.
For an example, an \emph{AR(1)} process can be identified by a exponentially-decaying ACF, with a single spike at the first lag in its PACF. Higher order models exhibit similar decay in their ACF and have a sharp cutoff in their PACF at the lag corresponding to the maximum order of the process.

\subsubsection{Moving-Average (MA) Process \label{sec:ma}}

When a variable is correlated with previous error terms in the
series, the process can be considered a moving-average of random shocks that
occurred in the recent past.  The errors commonly designated 
`innovations' in the econometrics and
statistics literature.  While \emph{AR} models are influenced by
previous \emph{values} of the series, \emph{MA} models trace the
influence of previous \emph{random innovations} that perturbed the
system.  Here the random noise term is intrinsically tied to the evolution of the
series, and is not an independent and separate effect;  each new random value adds
information to the series.  An \emph{MA(q)} model is described by
\begin{equation}
  \label{eq:ma}
  x_t = \epsilon_t + \theta_1 \epsilon_{t-1} + \theta_2 \epsilon_{t-2} + \ldots + \theta_q \epsilon_{t-q} 
\end{equation}
where $\epsilon_t$ is the error term for the $t$-th time point, $\theta_i$ is the coefficient for each lagged error term up to order $q$. 

Similarly to \emph{AR} processes, \emph{MA} models can also be identified through their (P)ACF, but with reversed behavior. A \emph{MA(1)} process has a exponentially-decaying PACF, with a sharp ACF cutoff at the lag corresponding to the maximum order of the process. Fitting \emph{MA} models is more complicated than fitting an \emph{AR} model, since here the covariates are a value that is not directly observed---namely $\epsilon_i$, the error terms. The parameters are typically calculated through an iterative procedure.

\subsubsection{Integrated Process}

In the presence of nonstationarity, stationarity can often be approximated 
through the differencing operation discussed in \S\ref{sec:acf}.  A
series is created by taking the point-to-point
difference in values, which may then be modeled as a stationary
\emph{ARMA} process. To recreate the original series and undoing the
differencing, the series is ``integrated'' back. The order of an integrated process refers to the number of differences needed to achieve stationarity for a given series. Often, a single difference is sufficient to achieve approximate stationarity.

A pure integrated process (that is, without any \emph{ARMA} components) can be described by
\begin{equation}
\label{eq:diff}
  (1-B)^d x_t = \epsilon_t
\end{equation}
where $d$ is the order of differencing and $B$ is the backshift operator defined by $Bx_t = x_{t-1}$. When $d$ is a positive integer, we referred to it as an \textit{integrated} process; when $d$ is allowed to take non-integer values, the series is \textit{fractionally integrated}.  Equation~(\ref{eq:diff1}) corresponds to the special case where $d=1$ in the more general description presented here. When a process is not stationary, significant correlation can be observed in the ACF up to very high lags.

As variable astrophysical processes like magnetic activity in stars or disk accretion in quasars are not restricted to following a deterministic functional form $X(t) = f(t) + \epsilon$ over extended times, we must be prepared to handle stochastic nonstationary behaviors such as random walks. A large class of such behaviors are difference stationary and can thus be readily treated within the ARIMA generalization of the ARMA model.

\subsubsection{ARIMA and ARFIMA}
\label{sec:arima}

The three components described can be incorporated into one
equation to jointly model more complex processes. ARIMA models
combine equations~(\ref{eq:ar})\nobreakdash-(\ref{eq:diff}) into a
single regression procedure where the $\theta$ and $\phi$ coefficients
are simultaneously inferred for the entire time series, and $d$ is either 
determined by the user or estimated separately.  The ARIMA model 
can be written as 
\begin{equation}
  \label{eq:arima}
  (1-B)^d x_t = \sum_{i=1}^p \phi_i x_{t-i} + \sum_{j=1}^q \theta_j \epsilon_{t-j} + \epsilon_t
\end{equation}

Generally, the best fit parameters for ARIMA-type models are determined using maximum
likelihood estimation. The calculation is made for a range of orders
$p$ and $q$, and the optimal model is chosen based on some
quantitative measure, such as the Akaike Information Criterion (AIC).

A variant of interest in astrophysics is when $d$ is a fraction rather than an integer; the process is called `fractionally integrated' ARMA, abbreviated ARFIMA or FARIMA \citep{Palma2007}.  When $0<d<0.5$, the result is a stationary long-memory autocorrelation with a power law behavior in both the autocorrelation function, $ACF(k) \propto k^{2d-1}$ for lag $k$, and in the Fourier spectral density, $f(\nu) \propto | \nu |^{-2d}$ for frequency $\nu$.  The process is nonstationary for $d>0.5$.  The ARFIMA parameter $d$ is arithmetically connected to $\alpha$ in the physicists' $1/f^\alpha$ `red noise' component where $\alpha = 2d$, and is connected to the econometrician's Hurst parameter, $H=d+0.5$.  

A mathematically profound basis for the effectiveness of ARMA-type models is the Wold Decomposition Theorem, which guarantees that any time series can be decomposed into the deterministic part plus an infinite sum of the innovations~\citep{Wold1938}.  An ARIMA$(p,d,q)$ model is a parsimonious approximation to this decomposition.

Figure~\ref{fig:lc-arma} shows how ARIMA-type modeling can reduce autocorrelation and noise in a Kepler light curve that the differencing operator alone (Figure~\ref{fig:lc-diff}) did not remove.   We have found that ARIMA and ARFIMA are highly effective in reducing non-planetary stellar variability for a large fraction of Kepler stars  \citep{Caceres2019b}.   This is consistent with the widespread success of ARIMA models for modeling and forecasting a wide range of time series on other fields.   ARIMA models are applied to model road accidents, stock prices, wind and weather fluctuations, solar irradiance, and innumerable other stochastic processes in human and physical systems.  ARFIMA models are less commonly used, but have been applied to model inflation and other macroeconomics measures, erratic heart beats in cardiology, and a variety of forecasting situations.

\section{Transit Comb Filter and Periodogram}
\label{sec:tcf}

\subsection{Motivation}

After investigating the effectiveness of ARIMA-type modeling to
characterize and remove unwanted stellar variability, we now turn to the search
for faint planetary transit signals in the model residuals. Transits
have the crucial property of strict periodicity, so that spectral
models that quantify intensity variations as a function of frequency
(or equivalently, period) will concentrate the transit signal. Most
transit detection techniques use periodograms produced by
quasi-Fourier procedures such as the Lomb-Scargle~\citep{Scargle1982}
or matched filters such as the Box-fitting Least Squares (BLS)
algorithm~\citep{Kovacs2002}.

The BLS approach fits a periodic series of box-shaped transits testing a number
of different periods, phases, and transit durations. This approach can be
equated to using a matched filter, where a signal is processed with a
filter of the exact shape expected (although the box is only an
approximation of a transit's shape).  For a signal in Gaussian noise, the matched filter is
equivalent to the maximum likelihood estimator. Thus we follow suit by
designing a new matched filter algorithm to optimize the search for
the modified transit signal.

\begin{figure}[t]
  \begin{center}
    \includegraphics[width=\textwidth]{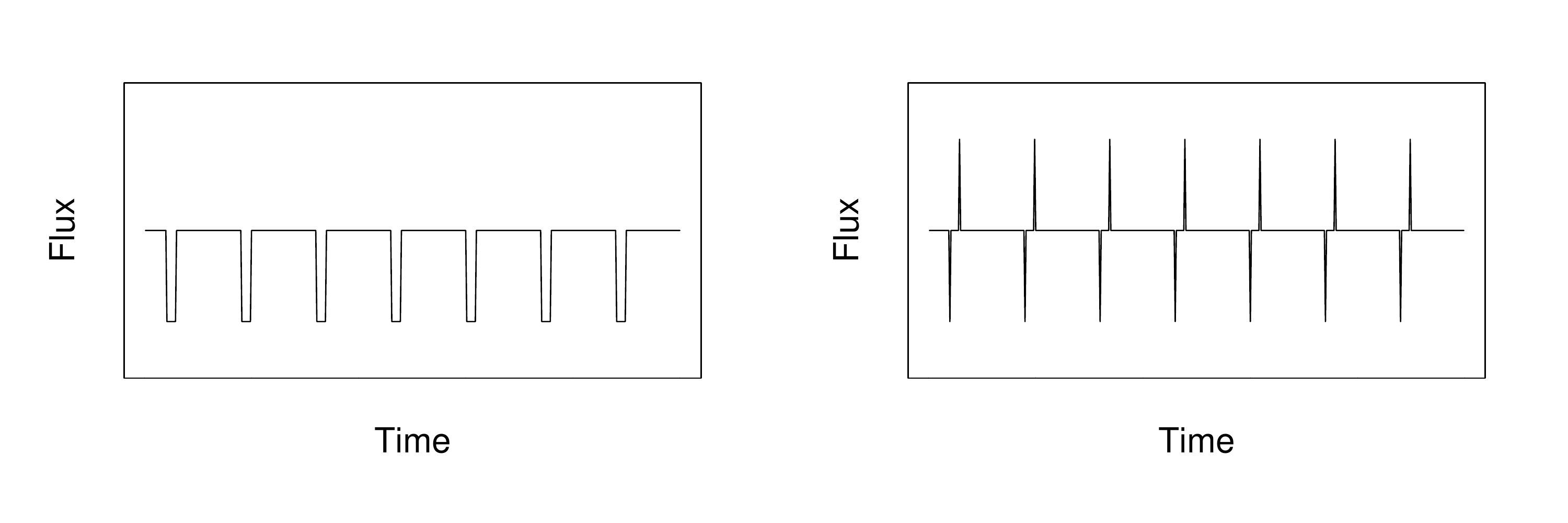}
  % \plotone{figures/diff-box}
  \caption{Schematic of box-shaped transits (\emph{left}) and the effects of differencing (\emph{right}).  A similar effect can occur from autoregressive modeling even without differencing.\label{fig:diff-box}}
  \end{center}
\end{figure}
\begin{figure}[t]
  \begin{center}
    \includegraphics[width=\textwidth]{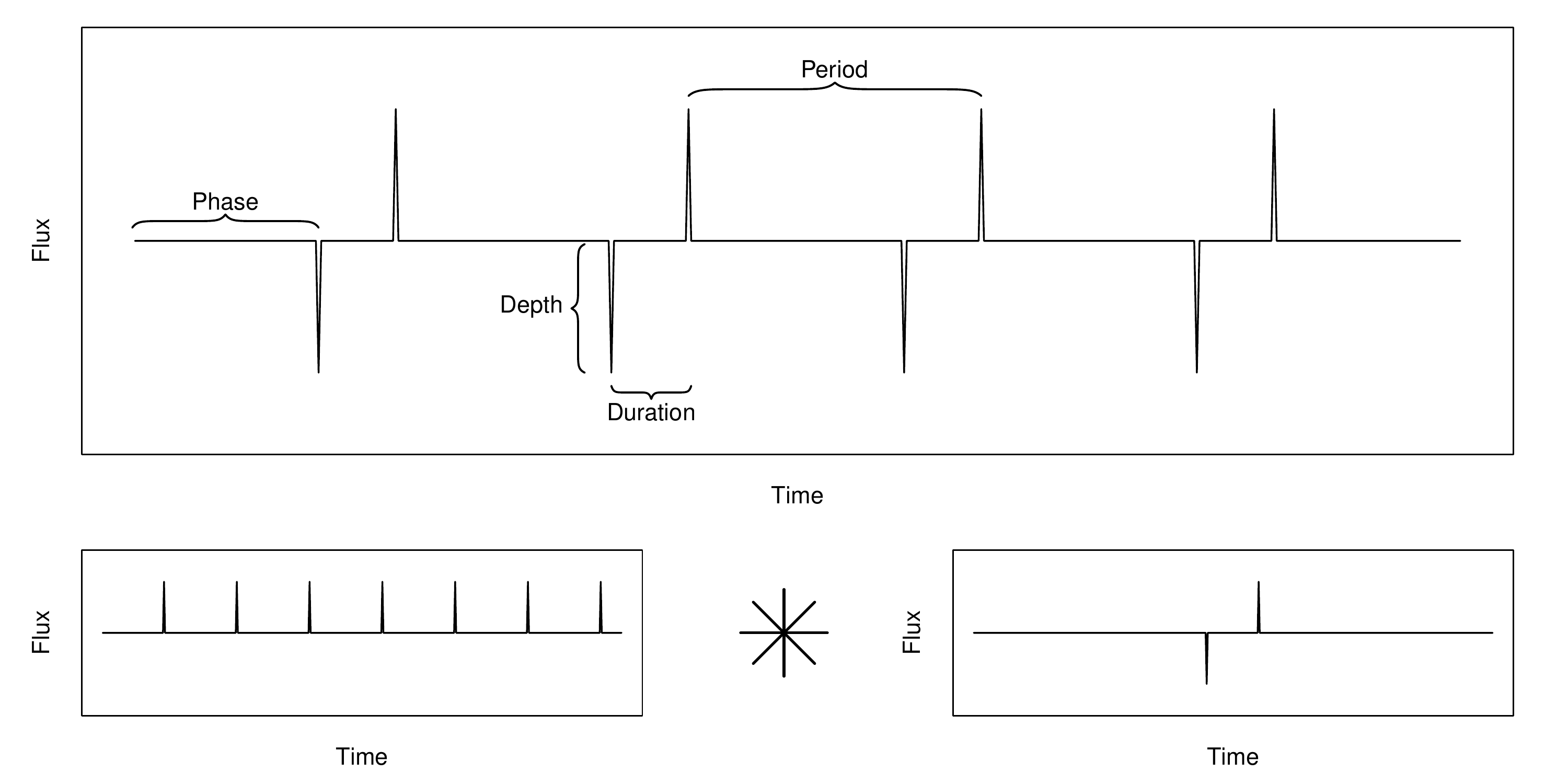}
  \caption{Schematic of the Transit Comb Filter.
\label{fig:tcf-sketch}}
  \end{center}
\end{figure}

\begin{figure}[t]
  \begin{center}
    \includegraphics[width=\textwidth]{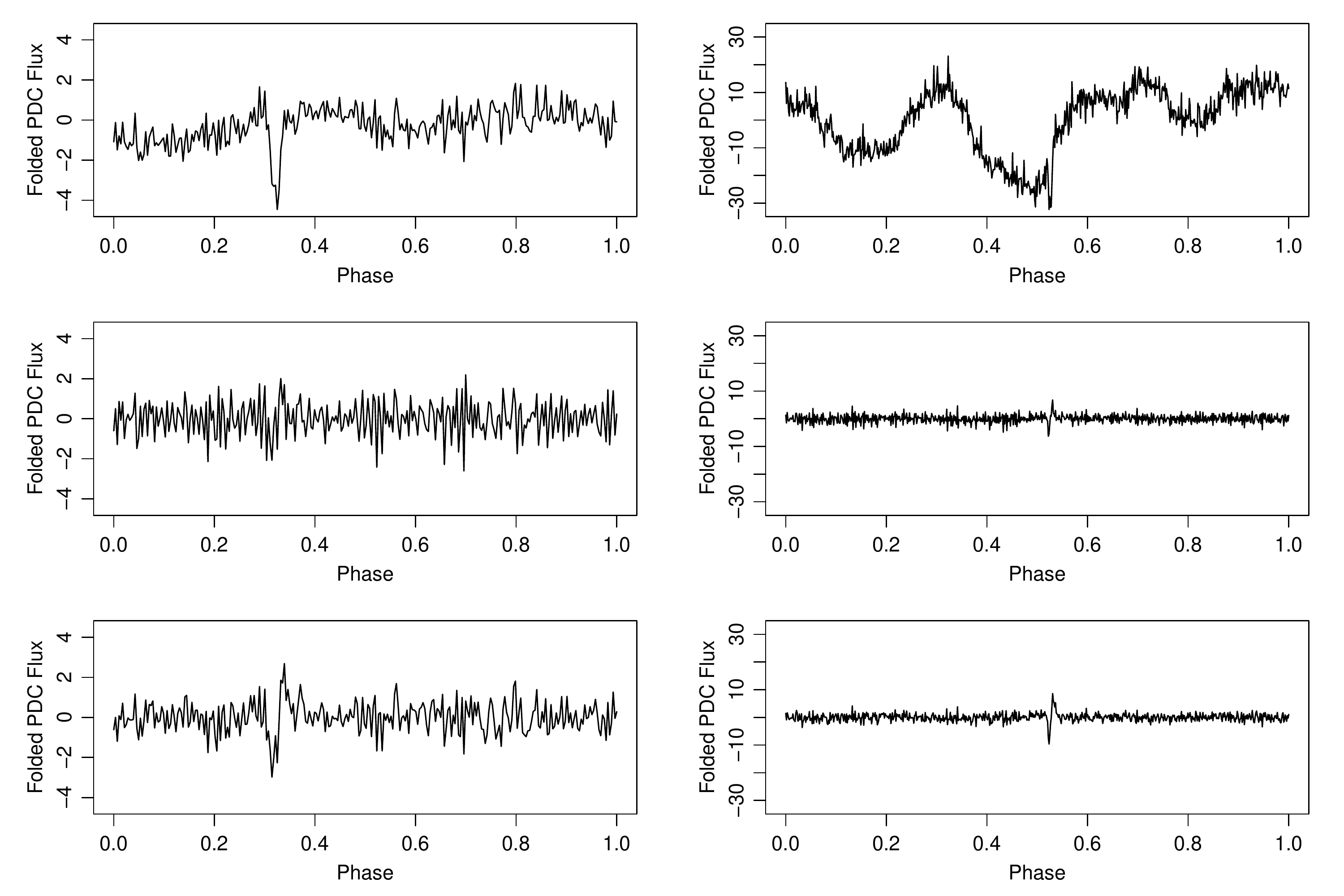}
  \caption{light curves of KIC 001724719 (left) and  KIC 010024701 (right) folded at the planetary transit period  at three stages of ARPS analysis: original light curve (top), after differencing (middle), and residuals from the ARIMA model (bottom).  
\label{fig:fold-ex}}
  \end{center}
\end{figure}

While the BLS algorithm could be applied to autoregressive residuals, its
performance will be degraded due to the differencing operation
described in \S\ref{sec:diff}. 
While ARIMA models are quite effective at reducing the noise inherent in
stellar light curves (\S\ref{sec:arima}), this gain has a cost. Differencing and 
autoregressive modeling affects the shape of the transit signal both in the 
case of a toy-model box transit 
(Figure~\ref{fig:diff-box}) and for real \textit{Kepler} stars (Figure~\ref{fig:fold-ex}). 
The periodic box-shaped dip in a stellar
light curve due to a planetary transit is transformed by the
differencing operator into a distinctive periodic double-spike
pattern.  The first spike is a decrease in flux due to the ingress of
the planet, and the second spike is an increase in flux due to the
egress of the planet.  During the intervening duration of the transit,
the flux is constant so the differenced time series returns to
zero. Additionally, random noise will be superposed on these
features.  

In order to best search for this modified
signal expected in the stellar light curve residuals after ARIMA
modeling, we devise a filtering algorithm to match its shape that we call the Transit Comb Filter (TCF), due
to its comb-like shape, and use it to create a new periodogram.

\subsection{Matched Filter and Parameter Estimation} \label{sec:TCFalgo}

The schematic in Figure~\ref{fig:tcf-sketch} shows the parameters of the transit. As shown in the lower panels, the shape of the periodic signal is a periodic sequence of Dirac delta functions denoted $\Shah(t)$, known as a Shah function in electrical engineering and as a Dirac comb in mathematics, convolved with a down-up double spike.  We call the resulting sequence a Transit Comb.  

Consider the simplified approximation where the evenly-spaced time series observed by \textit{Kepler} is a combination of a transit component and a white noise component,
\begin{equation}
  \label{eq:obs}
  x_t = s_t + \epsilon_t
\end{equation}
where, $x_t$ is the data (in ARPS, these are the residuals after ARIMA modeling), $s_t$ is the planetary transit signal shown in Figure~\ref{fig:tcf-sketch}, and $\epsilon \sim N(0,\sigma^2)$ is homoscedastic white noise with zero-mean and variance $\sigma^2$. We assume the residuals are approximately distributed as white noise, since after ``whitening'' the signal using the ARIMA models presented in \S\ref{sec:ar-models}, most of the autocorrelated behavior should have been removed. 

Let the filter $f_t$ have a known shape parameterized by period $P$, phase $\phi$, and duration $d$. Factoring out the transit depth $A$, any realization of the signal $s_t$ corresponds to a scaled copy of $f$ and is given by
\begin{equation}
  s_t = A \cdot f_t(P,\phi,d)
\end{equation}
We can estimate the optimal parameters $(A^*,P^*,\phi^*,d^*)$ of $s_t$ by minimizing the squared residuals between $s_t$ and the observed data, $x_t$. Thus
\begin{equation}
  \label{eq:mse}
  A^*,P^*,\phi^*,d^* = \operatorname*{argmin}_{A,P,\phi,d} \sum_t (x_t - s_t)^2 = \operatorname*{argmin}_{A,P,\phi,d} \sum_t (x_t-A \cdot f_{t}(P,\phi,d))^2.
\end{equation}
Expanding the square and separating the summations (omitting the arguments of $f_t$ to simplify the notation) gives
\begin{equation}
\label{eq:mse-expanded}
  A^*,P^*,\phi^*,d^* = \operatorname*{argmin}_{A,P,\phi,d} \left[ \sum_t (x_t)^2
                                             + A^2 \sum_t f^2_{t}
                                             - 2 A \sum x_t f_{t}\right]
\end{equation}
The first term in~(\ref{eq:mse-expanded}) corresponds to the observed data and can be ignored in the minimization since it is constant with respect to the transit parameters.
Reversing the sign to turn it into a maximization problem and simplifying gives
\begin{equation}
  \label{eq:mse-simplified}
  A^*,P^*,\phi^*,d^* = \operatorname*{argmax}_{A,P,\phi,d} \left[ 2 A \sum x_t f_{t} - A^2 \sum_t f^2_{t} \right]
\end{equation}
An analytic solution for $A^*$ as a function of the remaining parameters can be obtained from~(\ref{eq:mse-simplified}) by taking its derivative with respect to $A$ and setting it equal to zero, yielding
\begin{equation}
  \label{eq:depth}
  A^*(P,\phi,d) = \frac{\sum_t x_t f_{t}}{\sum_t f_t^2}.
\end{equation}
The optimal parameter estimates are are then obtained by maximizing 
\begin{equation}
  \label{eq:mse-final}
  P^*,\phi^*,d^* = \operatorname*{argmax}_{P,\phi,d} \frac{(\sum_t x_t f_{t})^2}{\sum_t f^2_{t}}.
\end{equation}

We now define the shape of the filter in terms of measurable quantities.  For the denominator of equation~(\ref{eq:mse-final}),  we first recognize that the number of transits $k$ in a time series has $N$ data points is 
\begin{equation}
  \label{eq:ntransits}
  k = \left\lfloor \frac{N - \phi}{P} \right\rfloor + 1
\end{equation}
where $\lfloor x \rfloor$ is the largest integer less than or equal to $x$.  The filter $f_t$ has a value of 1 or $-1$ at the location of each ``comb tooth'' (parametrized by $P^*$, $\phi^*$, and $d^*$), and equals zero everywhere else. Therefore $f_t^2 = |f_t|$ which has a value of 1 at each spike and zero elsewhere. Since there are $k$ transits with two spikes each\footnote{This value is an approximation since it is possible for a planet to be mid-transit when observations began or ended. But a single missing spike should not significantly affect our results since we require a candidate signal to have multiple transits.}, 
then $\sum |f_t| = 2k$. 

The numerator of equation~(\ref{eq:mse-final}) can also be simplified by noting that the product $x_t \cdot f_t$ equals $-x_t$ when $t$ corresponds to a downward spike (ingress), and $x_t$ when it corresponds to an upwards spike (egress). The ingress spike occurs at times $t$ that are multiples of the period plus a phase shift, and the egress spikes occurs at the same value plus an additional time shift due to the duration. Therefore the updated equation is
\begin{equation}
  \label{eq:tcf}
  P^*,\phi^*,d^* = \operatorname*{argmax}_{P,\phi,d} \frac{\left( \sum_{i = 1}^k x_{(i-1)P+\phi+d} - x_{(i-1)P+\phi} \right)^2}{2k}
\end{equation}

This formulation greatly reduces the number of computations required, since it omits all other values of $t$ that  evaluate to zero. The optimal values at which equation~(\ref{eq:tcf}) is maximized then correspond to the best estimate for the transit parameters.

We now have the ingredients for constructing a periodogram based on this least-squares search for periodic double spikes by taking a sequence of periods and for each maximizing~(\ref{eq:tcf}) to find $A^*$ and $\phi^*$. The pseudocode for the TCF implementation is shown in Figure~\ref{algo:tcf}.   The optimal value of~(\ref{eq:tcf}) gives a measure of the signal strength or ``power'' at the given period, and these values over all periods evaluated allows us to create a periodogram. The `best' period $P^*$ can reasonably be chosen to give the maximum power in the periodogram, or the maximum signal-to-noise ratio (SNR) with respect to local noise in the periodogram.  

\begin{figure}[t]
  \begin{center}
    \includegraphics[width=\textwidth]{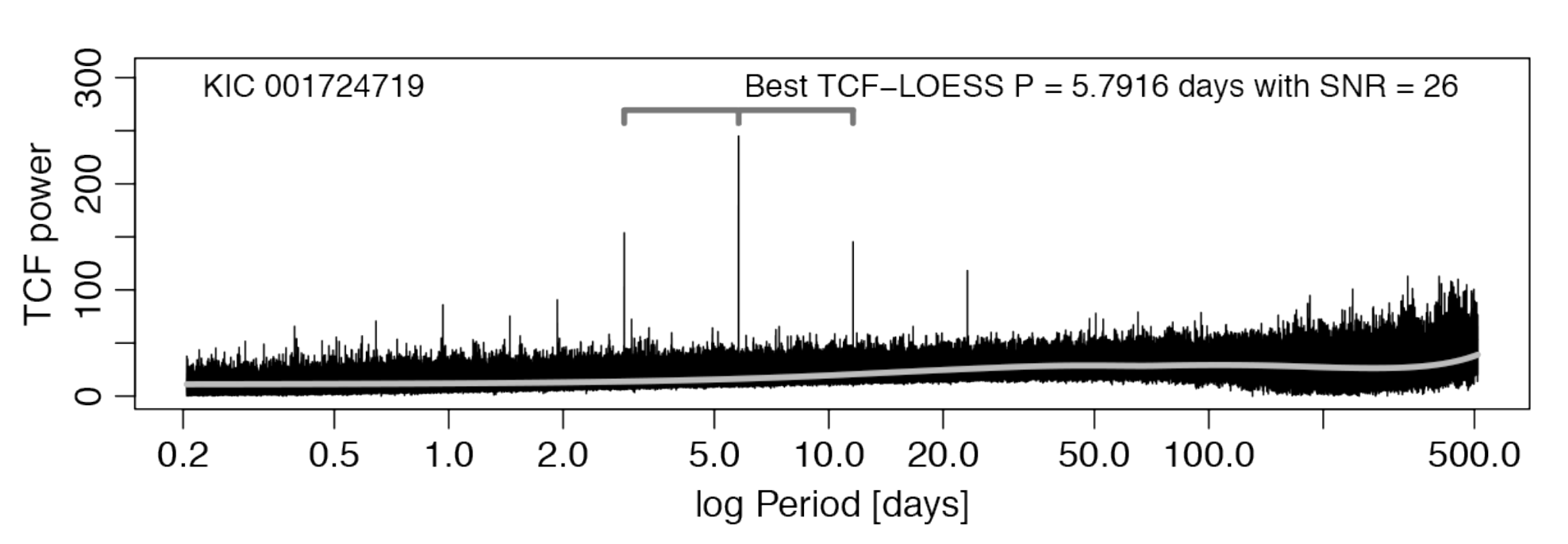}
    \includegraphics[width=\textwidth]{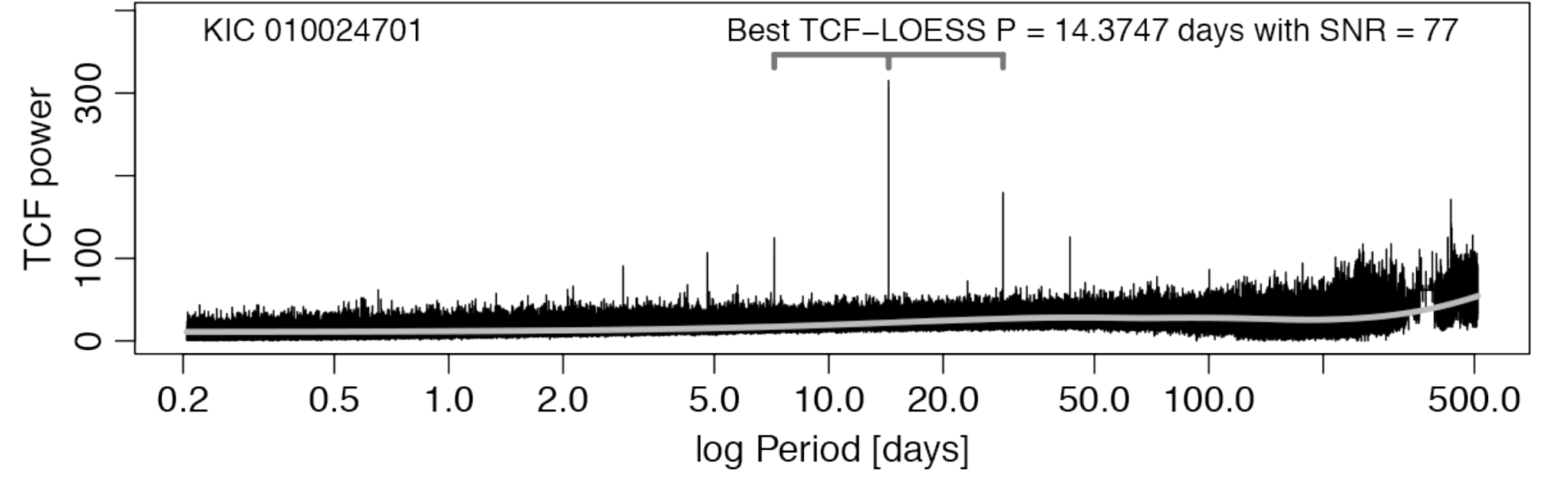}
    \caption{Transit Comb Filter periodograms for ARIMA residuals of the two Kepler stars showing matched filter power for $\sim$500,000 periods ranging from 0.2 to 500 days.  The gray curve shows the LOESS smoothed estimator for the median of the TCF power.  The periodogram peak is marked with 1/2-times and 2-times harmonics.  The signal-to-noise ratio is calculated with respect to noise in a window around the peak.} \label{fig:TCF_KIC}
  \end{center}
\end{figure}

The application to the ARIMA residuals of the two Kepler stars is shown in Figure~\ref{fig:TCF_KIC}.  In each case, we see a clear `best' period with factor-of-two harmonics as expected from a true periodicity.  Here the same period has maximum power and maximum SNR in the periodogram, but this is not necessarily the case.  

\begin{figure}[t]
\algblock[Name]{Start}{End}
\algblock[Name]{Algorithm}{End}
\begin{algorithmic}
\Algorithm {\bf ~1:  Transit Comb Filter pseudocode } \\  \label{TCFalgorithm}
\Require {Equally-spaced light curve, periods (in units of cadence) \\
\hspace{-11pt}{\bf Result:}  TCF periodogram, transit parameters at each period \\ }
\Start 
\For{{\it each period}} 
	\For{{\it each phase}} \\
	\qquad\qquad	// Dot product between Shah filter and light curve \\
	\qquad\qquad       $sum_{phase} \leftarrow \displaystyle \sum^N_{i=1} \Shah_i x_i$\;
	\EndFor	
	\For{{\it each duration}} \\
	 \qquad\qquad	// Average between egress and ingress \\
         \qquad\qquad     $depth_{phase} \leftarrow \displaystyle \frac{(sum_{phase + duration} - sum_{phase})}{2N}$\;  \\
         \qquad\qquad     $power_{phase} \leftarrow \displaystyle \frac{(sum_{phase + duration} - sum_{phase})^2}{2N}$\;
         \EndFor
         
         Record $power$, $depth$, $duration$, and $phase$ corresponding to the maximum \\ \qquad value of $power_{phase}$, for current $period$\;
\EndFor
\End
\End
\end{algorithmic}
\caption{{\bf Algorithm: Transit Comb Filter pseudocode} } \label{algo:tcf}
\end{figure}

Careful consideration is needed for the choice of  periods to be examined.  While it is common practice to sample periods for a periodogram in constant-frequency steps, \citet{Ofir14} notes that this approach may not be optimal for transit detection when using an algorithm like BLS due to unnecessary oversampling of short periods or damaging undersampling of long periods. We adopt Ofir's recommendation to use $\Delta f \simeq q / S$ where $q$ is the duty cycle (i.e., transit duration in fractional phase) of the signal and $S$ is the full span of the time series in candences.  Since the TCF focuses on the ingress/egress spikes, the duty cycle would be $q \simeq 1/P$ where $P$ is the period, giving a corresponding relation 

\begin{equation} \label{TCF_periods1.eqn}
\Delta f \simeq \frac{f}{S}.
\end{equation}

We can arrive to a similar conclusion estimating the error of a transit's ingress time when using an incorrect period. If one assumes that the transit's ingress and the measurement occur instantly, and if the period is incorrect by $\Delta P$, then each transit is delayed by that amount leading to the n-th transit being off by $\Delta P \times n$. A signal with period $P$ observed over a span of time $S$ has at most $n=S/P$ transits so that $\Delta P \simeq P/S$.  Inverting this gives the relation in equation (\ref{TCF_periods1.eqn}).  An equivalent formulation is to use periods 

\begin{equation} \label{TCF_periods2.eqn}
P_k = P_0 \bigg(1+\frac{1}{S} \bigg)^k
\end{equation}

\noindent where $P_0$ is a chosen minimum period and $k$ is a sequence of integers noting the number of points in the periodogram.  

A few approximations are made for equations~\ref{TCF_periods1.eqn}-\ref{TCF_periods2.eqn}. Realistic transits do not align exactly with the instant of the first observation, so the $\sim 1$ cadence bin shift over the span of the time series means that later transits may shift into the next cadence bin. Furthermore, ingresses have a finite duration and the observation is integrated over a period of time, which can also lead to a one-cadence shift.  To reduce these effects, the "teeth" of the Transit Comb Filter can be given widths larger than a single cadence.

\subsection{Estimating the significance of TCF periodogram peaks}
\label{sec:arbox}

As with other periodograms, evaluation of the False Alarm Rate of peaks in the TCF periodogram is difficult.  Except for unrealistic ideal situations, analytic calculations of periodogram significance levels are unreliable for Lomb-Scargle periodograms of irregularly spaced data~\citep{Koen1990, Vaughan2016, Vanderplas2018}, and even for Fourier periodograms of regularly spaced data~\citep{Percival1993}.

We have found in practice that the distribution of a TCF periodogram powers in the absence of a periodic signal is highly non-Gaussian and with variable mean.  In Kepler periodograms, a nonlinear trend is seen with medians rising as period increases.  We remove this trend with a smooth local regression fit to the medians of the TCF periodogram using the well-established LOESS algorithm (Cleveland 1981).  A similar behavior is found in Box Least Squares periodograms of Kepler light curves as discussed by Ofir (2014) who also recommends nonparametric detrending of the periodogram.  After trends are removed by subtracting the LOESS curve, we record the highest-power peaks in the periodogram and estimate a local signal-to-noise ratio (SNR) within a window of nearby period values.   This is illustrated for the two Kepler stars in Figure~\ref{fig:TCF_KIC}.

However, we can assist with evaluating periodogram peak significance by estimating the marginal likelihood of the transit depth parameter $A$, and thereby estimate its signal-to-noise ratio.  Section~\ref{sec:TCFalgo} noted that $A$ can be factored out of the filter and simply calculated as a function of the other parameters.  This is possible because, given parameters $p$, $\phi$, and $d$, determining $A$ is a simple linear problem; the period, phase, and duration fully describe the location of each ``box'' and all that remains to calculate is its respective amplitude. 

We therefore add an additional step of fitting an ARIMA-type model that includes a simple box-shaped model of the transit corresponding to the best TCF period.  The incorporation of covariates in ARMA-type modeling is common in econometrics.  Our situation is similar to the econometric problem of modeling retail sales with both stochastic autoregressive characteristics and a deterministic weekly cycle; here the period (7 days) and phase (weekday $vs.$ weekend) is known but the amplitude of the cyclical component is unknown.  The statistical model is often called ARIMAX for ARIMA with `explanatory' or `exogeneous' variables, or `dynamic regression' \citep{Hyndman2014, Box2015}.  

This idea is formalized by defining an indicator function, $I(t)$, which equals 1 when in transit (as determined by the given $p$, $\phi$, and $d$) and 0 otherwise. These values are obtained from the `best' period in the TCF periodogram.   To regress the observed data on this indicator function, the corresponding coefficient of this linear regression corresponds to the depth of the box. That is
\begin{equation}
  \label{eq:ols}
  x(t) = \beta_0 + \beta_1 I(t) + \epsilon_t
\end{equation}
where $\beta_0$ fits the global mean, $\beta_1$ is the coefficient of the indicator function, and  $\epsilon \sim N(0,\sigma^2)$ is the error term. The $\beta_1$ coefficient is the mean offset while in transit (that is, the transit depth).  

This deterministic regression is combined with the stochastic ARMA model by dropping the assumption that the error term of equation~(\ref{eq:ols}) is white Gaussian noise, but instead is an autocorrelated  ARIMA process. The hierarchical regression model with ARMA errors (temporarily ignoring the differencing operator to simplify the notation) then looks like
\begin{align}
  \label{eq:ar-err}
  x(t) &= \beta_0 + \beta_1 I(t) + \eta_t \\
  \eta_t &= \sum_{i=1}^p \phi_i \eta_{t-i} + \sum_{j=1}^q \theta_i \epsilon_{t-i} + \epsilon_t \nonumber
\end{align}
where $\epsilon_t \sim N(0,\sigma^2)$ is the white Gaussian noise term.  We now compute the maximum likelihood values in the model (\ref{eq:ar-err}), estimate the value and uncertainty of the $\beta_1$ parameter from the linear regression model, and compute the depth amplitude signal-to-noise ratio

\begin{equation}
\label{eq:a-snr}
SNR(A) \simeq \hat{\beta_1} / \widehat{\sigma_{\beta_1}}. 
\end{equation}
Note that signal-to-noise ratio shown in Figure~\ref{fig:TCF_KIC} is different from that of equation (\ref{eq:ar-err}); the former measures significance of the TCF power before a periodicity has been identified, while the latter measures the significance of the transit depth after a periodicity is assumed to be present.

While this ARIMAX approach may seem to be redundant to the earlier ARIMA approach, we find it can be very valuable in assisting the evaluation of the significance of a TCF periodogram peak.   In our application to the Kepler 4-year dataset, the ARIMAX $SNR(A)$ from equation (\ref{eq:a-snr}) proved to be the most important `feature' in the machine learning classifier for identifying planetary candidates and discriminating them from False Positives like blended eclipsing binary stars. 

\subsection{Comparison of TCF and BLS periodicity search}

\begin{figure}[t]
  \begin{center}
    \includegraphics[width=0.95\textwidth]{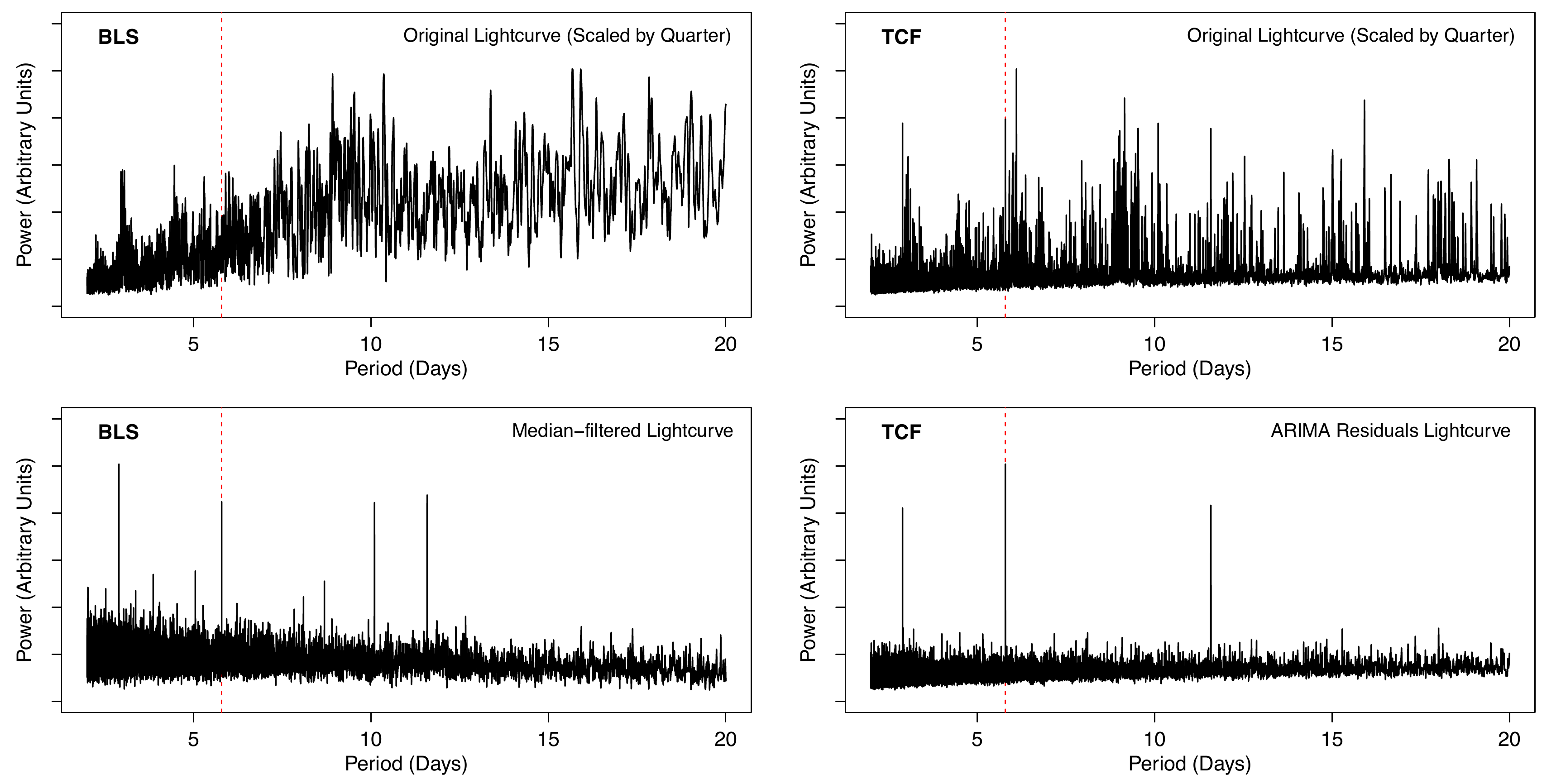}
  \caption{Comparison between BLS (left panels) and TCF (right panels) periodograms for KIC 001724719. The top panels show each algorithm applied directly to the original light curve. The bottom-left panel shows the BLS periodogram of a median-filtered light curve using a 12-hour window. The bottom-right panel shows the TCF periodogram applied to ARIMA residuals. The dashed red line shows the known planetary period. \label{fig:bls_v_tcf_001724719}}
  \end{center}
\end{figure}
\begin{figure}
  \begin{center}
    \includegraphics[width=0.95\textwidth]{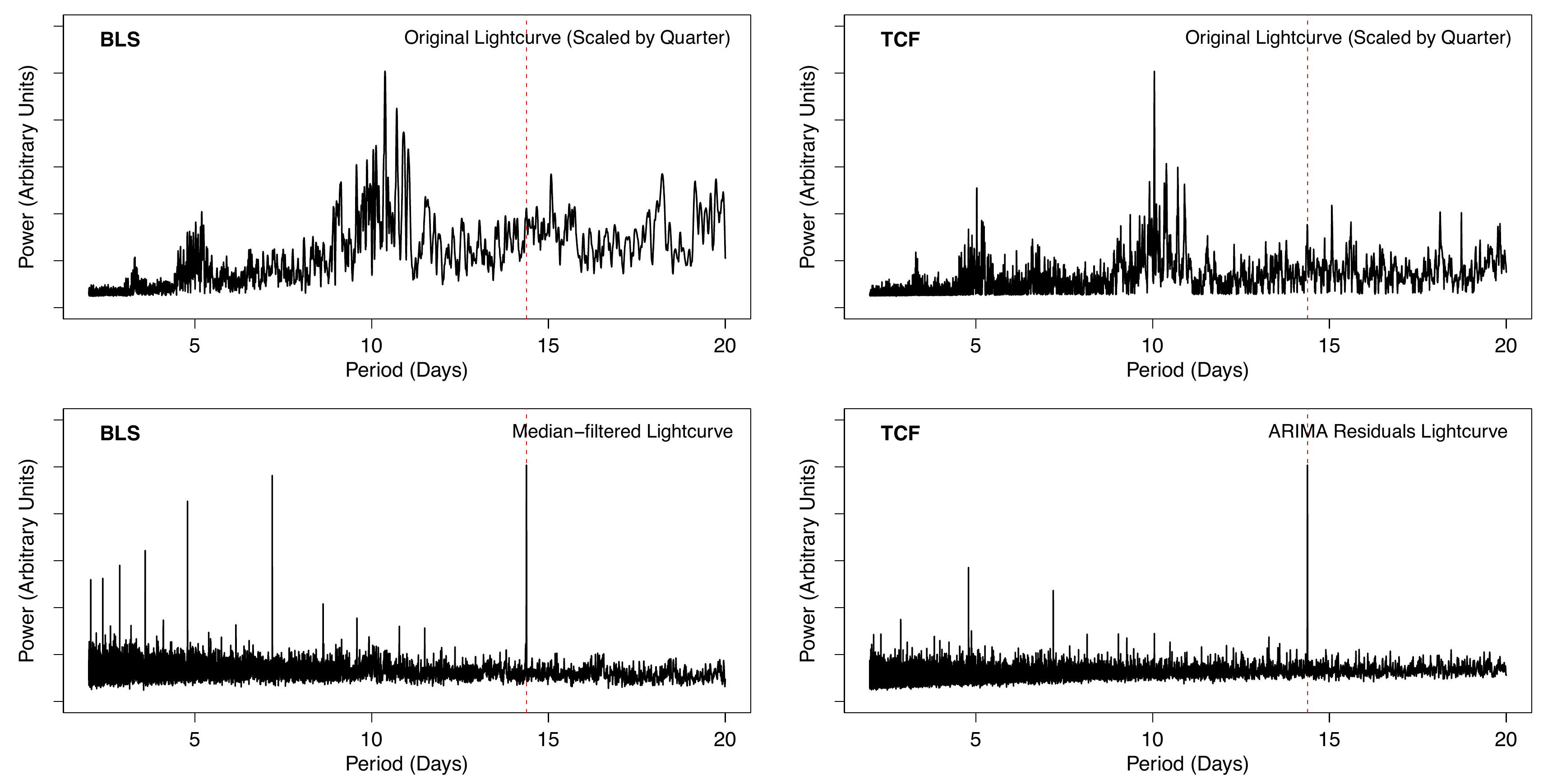}
     \caption{ Same as Figure~\ref{fig:bls_v_tcf_001724719} for KIC 010024701.} \label{fig:bls_v_tcf_010024701}
  \end{center}
  \vspace{20pt}
\end{figure}

Figure~\ref{fig:TCF_KIC} showed how the TCF algorithm was capable of recovering the transit signal from ARIMA residuals for both of our example stars. To illustrate how this compares with other approaches, we now show a simple application of the BLS algorithm to both stars. These results are summarized in Figures~\ref{fig:bls_v_tcf_001724719}~and~\ref{fig:bls_v_tcf_010024701}. In each figure, the top panels show BLS and TCF applied to the original PDC light curve (with each quarter scaled to zero-median). The bottom panels show the BLS periodogram applied to a median-filtered light curve with a 12-hour window, and the TCF periodogram applied to the ARIMA residuals.  We caution that this comparison of TCF and BLS performance is very incomplete.   The mathematical treatments are different, the periodograms measure different quantities, and the sensitivities of the methods may differ when applied to time series with different cadences, stellar noise characteristics, planetary transit depths and durations. 

Neither algorithm detects a significant transit signal in the original light curve; this emphasizes the importance of some type of filtering even when a light curve may not exhibit strong variations. After reducing stellar variability, the true planetary periods are clearly seen in both the BLS and TCF periodograms, but the BLS periodogram shows stronger harmonic structure.  The harmonic structure can help to confirm the existence of a periodicity, but may also confuse identification of the true peak when the 1/2-times or 2-times harmonics have comparable power.  The BLS periodograms also tend to have higher noise at shorter periods than than the corresponding TCF periodograms.   

BLS performance will be affected by different filtering procedures such as median filter, Fourier filter, wavelet decomposition, Gaussian Processes or other local regression.   There is no mathematical guidance regarding the choice of filter, although properties of the residuals  can be compared to improve the choice for a given star.   In most of these nonparametric procedures, a suitable bandwidth and/or kernel needs to be selected; the resulting BLS periodogram may be considerably affected by these choices.   Most astronomical researchers choose the bandwidth heuristically, while statisticians would recommend bandwidths chosen by cross-validation to minimize the global sum of bias-squared and variance \citep{Wasserman2007}.  However, for the purposes of periodic transit detection, the optimal value of this parameter may depend on the exact properties of period, depth, and duration which are unknown for yet-undiscovered objects. The 12-hour boxcar window used for the BLS analysis in Figures~\ref{fig:bls_v_tcf_001724719}~and~\ref{fig:bls_v_tcf_010024701} was selected by evaluating the periodogram using different smoothing windows. In practice, this may not be computationally viable when a large range of periods are being tested for many objects. 

A serious concern is whether TCF will have reduced sensitivity to transits compared to BLS because the modified Shah function template is matched only to ingress and egress spikes, while the BLS fit includes photometry during the transit as well as the ingress/egress events.  The strength of a transit depth in a light curve can be viewed as the product of the depth and $n_t$ where $n_t$ is the product of the number of transits observed during the observation span and the typical number of points per transit. For a box shape, $n_t$ scales with transit duration $d$, while the number of points per transit for TCF is always close to unity (where the exact values depend on the width of the ingress `teeth' chosen for the calculation) independent of duration.  We thus expect ARPS analysis to be most sensitive to short-duration transits where TCF inherits the advantages of ARIMA modeling while still retaining most of the transit signal.

The deterioration of TCF sensitivity is linked to period for the simple case of planets in circular orbits with zero inclination. The transit duration is then $d = R_* / v_{orb}$ where $v_{orb} = 2 \pi a / P$ where $R_*$ is the stellar radius and $a$ is the planet's semi-major axis.   As Kepler's Law requires $P^2 \propto a^3$, the transit duration is related to orbital period as $d \propto P^{1/3}$.  Therefore, for an algorithm like BLS, the reduction in $n_t$ for planets with longer periods is partially offset by the longer duration caused by the slower planetary orbital velocity.   Specifically, strength of the BLS periodogram peak will scale as $\sqrt{n_t} \propto \sqrt{d / P} \propto P^{-1/3}$.  In contrast, the TCF periodogram peak is independent of duration and scales as $\sqrt{n_t} \propto \sqrt{1 / P} \propto P^{-1/2}$.   Comparing a host star with identical planets at orbital periods 1:10:100 days, the BLS periodogram peak weakens as 1.00~:~0.46~:~0.22, while the TCF periodogram peak weakens as 1.00~:~0.32~:~0.10.  

We therefore expect TCF to exhibit its best sensitivity with respect to BLS at short periods like $P \simeq 1$ day with moderate deterioration (1 - 0.32 / 0.46 = 30\%) at 10 days and stronger deterioration (1 - 0.10/0.22 = 55\%) at 100 days.  The sensitivity loss does not apply to planets with significant ellipticity and/or inclination where the transit duration can be short even for long period orbits.  This calculation also does not evaluate the coefficient in front of these proportionalities;  the TCF periodogram peak could be higher or lower signal-to-noise than BLS periodogram peak at any given period for a given planet.  We do not know any reliable calculation of the relative sensitivities of the two methods for realistic light curves where complicated treatments (ARIMA modeling, nonparametric wavelet analysis, local regression, or other procedure) have been applied to reduce the aperiodic variability.  The possible presence of non-Gaussianity and autocorrelation in the residuals, possible removal of unknown portions of the periodic transit in earlier modeling steps, and complicated alias structure in the periodograms together make it difficult to estimate the signal-to-noise for any periodogram in realistic cases.  The cases shown in Figures~\ref{fig:bls_v_tcf_001724719}-\ref{fig:bls_v_tcf_010024701} indicate that TCF can perform well in comparison to BLS for these Kepler light curves with periods around $5-15$ days.  Its relative performance may be even better for shorter periods including UltraShort Period ($P \leq 1$ day) planets.

\section{Classification}
\label{sec:class}

\subsection{Learning Algorithms \& Feature Selection}

While a single periodogram can be investigated and assessed by eye, often we have many light curves that need to be evaluated, whether from surveys or simulated data. These large samples provide both a challenge and an opportunity. The large number of objects can make individual assessments prohibitively time consuming and raises the challenge of how to automatically select promising candidates for further exploration while reducing contamination from random statistical variations and from astronomical False Positives such as eclipsing binary light curves.  The ability to evaluate the performance of the approach on a wide variety of objects can enable us to make improved statistical statements on the significance of a given discovery. We review the use of learning and classification algorithms to address this issue, and discuss how to define automatic selection criteria for detection.  The exact procedure depends on the dataset under study, so our treatment here is schematic. 

Quite often, a single attribute such as periodogram power is used to determine whether the signal in a given object is significant enough for scientific interest. Section~\ref{sec:arbox} discussed two approaches: estimating the strength of a peak in the periodogram (either its spectral power or its signal-to-noise ratio relative to the local periodogram noise), and assessing the significance of the transit depth within a parametric ARIMAX-type model with both stochastic autoregressive and deterministic periodic components. 

But other properties or `features' of the data at different stages of ARPS processing can also assist in evaluate a possible transit.   One approach for analyzing time series data is to create summary variables characterizing relevant parts of the data \citep{Wang06}. Guiding principles have been developed on the choice of features for multivariate classification \citep[see an overview by][]{Guyon03}.  \citet{Armstrong2018} gives an example of feature engineering for transit planet detection.  Feature selection for classification of ensembles of astronomical light curves outside of exoplanetary detection is discussed by \citet{Richards11}, \citet{Graham13}, \citet{Rimoldini14}, \citet{Benavente17}, \citet{Pashchenko18}, \citet{Cabral18}, and others.  For the ARPS method, we take the approach of incorporating a variety of different measures summarizing both the original light curve as well as intermediate data products from the autoregressive model and the TCF periodogram. 

Features that can be used to inform the classifier can include: properties of the star like the magnitude and radius of the star; properties of the original,  differenced, and ARIMA residual  light curves such as IQR and Durbin-Watson statistic (a scalar measure of the ACF amplitude at lag=1); properties of the TCF periodogram peak such as period, signal-to-noise, and presence of harmonics; and properties of the folded light curve such as shape statistics, depth amplitude and duration, and comparison of even-vs.-odd events.  

Hard and soft classification are two common terms used to describe the output from a classification algorithm. The former applies when a specific class prediction is required, while the latter is used when probabilities for each star belonging in each category are calculated instead.

In order to apply supervised learning algorithms we require a ``labeled'' training set, where the expected output is known. This can take the form of data previously classified through other means, or simulations where the true values are known.  For exoplanetary studies, simulations can take the form of injections of artificial transits into real stellar light curves~\citep{Christiansen2016}.

\subsection{Random Forest}

The Random Forest classifier, an extension of classical decision tree classifiers involving sequential splits of the data based on critical values of different variables, is a powerful approach to  the selection of new planetary candidates \citep{Breiman2001}. The 'Autovetter' of the Kepler Team is based on Random Forest decision trees~\citep[][see also Mislis et al. 2016 and Armstrong et al. 2018]{McCauliff2015}.  \citep[In contrast, the Kepler Team 'Robovetter' procedure classifies transits with a thresholded univariate metric based on the shapes of folded light curves,][]{Thompson2015}.  Other multivariate machine learning approaches can be considered such as Support Vector Machines and neural networks including Deep Learning convolutional networks. 

A Random Forest is an ensemble algorithm made up of a multitude of decision trees \citep{Breiman2001}. Each tree generates many sequential, binary splits of the data based on a single variable in order to discriminate between the categories of the response variable. At each split (called a node), thresholds are tested for predictor variables and the threshold-variable combination which best optimizes a defined criterion is used for the split. Typical criteria for decision tree classification are to decrease Gini impurity or increase information gain.

Random Forests have important advantages over other classifiers for our problem.  Since each variable is tested independently and only the relative rank of the values, not their magnitudes, are used for splitting, decision trees -- and by extension Random Forests -- are not affected by differing units and scales (e.g. logarithmic transformations) of the input variables.  Unlike many other machine learning methods, no scale-dependent metric is used.  Input variables of potential utility can be proliferated because the method is robust to uninformative variables \citep{Cutler12}.  Dozens of features can be generated for the method to consider in its optimization of a classifier.   Random Forests  are computationally fast.  Random Forests provide outputs that assist in understanding the basis of the classification:  the relative importance of each variable is provided, and any single decision tree can by understood as a sequence of univariate decision rules.  In contrast, the output of many other classifiers (such as neural networks) are difficult to interpret.  

Decision trees procedures have weaknesses.  They are often biased towards variables with greater number of categories, or towards continuous variables instead of categorical ones~\citep{Breiman84, White94,  Loh2002, Strobl07}. Measures of variable importance can be biased in such cases \citep{Genuer10}.  Individual trees are often prone to overfitting, and may benefit from additional methods such as pruning~\citep{Quinlan1987}.  While Random Forests can maintain its predictive power when given highly-correlated predictors \citep{Strobl07}, it can be overwhelmed if many highly correlated variables lead to similar trees with similar splits \citep{Cutler12}.  Best performance is achieved when the ensemble's trees are not strongly correlated.

A 'forest' is obtained by training an ensemble of trees, each based on a bootstrap resampling of the original data (i.e., a random sample, with replacement, of entries of the data and of the same size as the original data). At each split only a random subset of the predictor variables are used. This allows variables that are slightly less important to still contribute to the decision making, and can also help reduce the bias towards variables with greater number of differing values. Using an ensemble of trees helps overcome some of the weaknesses of individual decision trees, reducing the variance of the global classifier at the expense of additional computational effort and a less intuitive underlying model.

\begin{figure}[t]
  \label{fig:rf_schematic}
  \begin{center}
    \includegraphics[width=5in,height=3.5in]{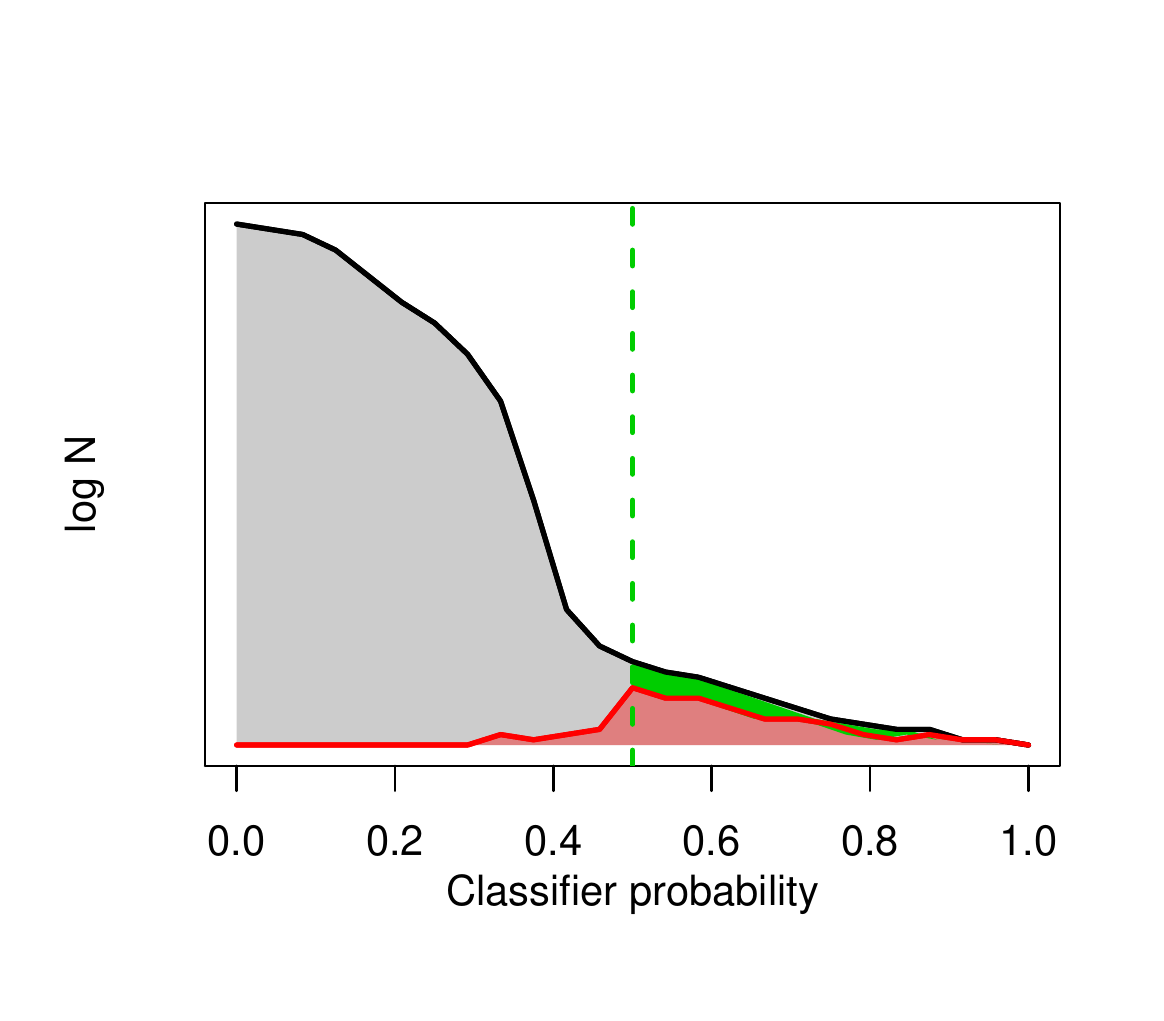}
    \caption{Schematic application of a Random Forest classifier to test data. Dashed green line: subjectively chosen threshold for selection of candidate planets.  Red region: Stars with confirmed planets from training set.  Green region: Stars with newly discovered candidate planets.  Grey region: Unselected stars.} \label{fig:RF_test}
  \end{center}
\end{figure}

The trees of the forest are combined into a single probabilistic result using bootstrap aggregation, also known as bagging~\citep{Breiman1996,Breiman2001}.  Typically, the majority vote of the trees in a Random Forest is used for hard classification, but the vote fractions represent soft classification and provide an estimate of the model's certainty about its decision. Figure~\ref{fig:RF_test} shows a schematic of the class probability estimated by a classifier to discover new candidate planets.  Known planetary candidates (red region) make up a small fraction of the full stellar population (grey region) and typically are given a higher score by the classifier. The remaining objects above a certain threshold which are not known candidates (green region) correspond to the discovery space of new candidates. The dashed green line corresponds to the cut-off set to maximize the recovery of known planets while minimizing false positive classifications.

The standard tunable parameters in a Random Forest are the total number of trees to grow and the number of features to test at each split \citep{Cutler12}. These two parameters allow generation of a Random Forest: the vote of many trees grown on bootstrap samples of the data \citep[`bagging',][]{Breiman96}, and random feature selection at each node \citep[`subspace method',][]{Ho98}.  It is possible to tweak other parameters, such as tree depth and minimum node splitting size, but this is not commonly done.  In standard implementations of Random Forests, trees are fully grown and branches are not pruned to reduce model complexity \citep{Quinlan1987}. In R's {\it randomForest}, the default procedure is to split the data as many times as necessary until each terminal node contains only a single data point. While it may appear unintuitive to grow such complex trees and thus overfit the data, this allows reduction of variance in the ensemble's prediction through bagging. 

 The probabilities estimated by this Random Forest model are uncalibrated. Although it has been argued that Random Forest predictions can approximate true posterior probability distributions \citep{Fan05}, this is a more challenging problem than the simpler case of developing a good discriminator. Furthermore, undersampling the data to address class imbalance during training affects the estimated probability of a model. The modified probabilities do not affect the rank-order of the estimates; this maintains the discriminatory power of the classifier, but affect the probability calibration. Methods do exist to recalibrate the estimates \citep{DalPozzolo15} and, more generally, to apply classification models to test sets that have different class distributions than the original training set \citep{Elkan01, Saerens02}. 

\subsection{Training Set Specification}

Many textbook cases of classification have unambiguous labels for the desired classes (such as the numbers `1', `2' and `3' in optical character recognition) and no difficulty constructing balanced training sets with similar number of objects in each class (such as `cat' and `dog' in photographs).   Both of these issues are problematic in machine learning applications to exoplanet transit detection.  

The first problem arises because the ensemble of stellar light curves without known planets is highly heterogeneous, ranging from quiet stars with no noticeable signals, to stars with quasi-periodic rotationally modulated starspots, eclipsing binaries, explosively flaring stars, and other variables.  In addition, some light curves may be subject to instrumental and observational effects that mimic some aspects of planetary transits with no relation to the star itself.  This wide range of light curve behaviors needs to be properly sampled in the `non-planet' training set to reduce the chance of the algorithm mistaking them for planetary transits. 

It is not obvious whether all types of `non-planet' should be grouped into a single class for the Random Forest algorithm, or whether better discrimination will be possible by considering several classes of `non-planet'.  Care must also be taken for which `non-planet' objects are included since any new transit discoveries must obviously come from systems that previously had no known planets. If a light curve containing a yet-undiscovered planet is used during training with a negative label, then the algorithm could learn that the signal does not originate from a planet and thus fail to discover it. Research groups involved in space-based \citep{McCauliff2015} and  ground-based \citep{Schanche18} transit surveys invest great effort in discriminating light curves that are convincingly true exoplanetary surveys from those with some periodic signals that arise from other astronomical causes, such as blended eclipsing binaries.  The judgment between `certified' planet candidates and False Positives can be difficult to make.  

The second problem is due to the unavoidable geometrical constraint that only a small fraction of planets will have orbital inclinations allowing transits to be detected.  This can be compounded by survey sensitivity; for example, a ground-based survey subject to noisy observational conditions may be sensitive only to hot Jupiter transits which are very rare.  The result is a highly imbalanced sample; there may be dozens, or even thousands, of stellar light curves without transits for each one with a transit.   Furthermore, the small size of the `planet' training set may poorly span the high-dimensional multivariate feature space, give rise to noisy and inaccurate classifiers.  

It is possible to down-sample the large class or up-sample the small class to construct artificially balanced training sets for classifier training.  The SMOTE algorithm, for example, is widely used for proliferating simulated objects in a small training set~\citep{Chawla2002}, and random sampling is used for pruning objects in a large training set.  However, it is difficult to overcome extreme class imbalances, as seen in exoplanetary surveys, with these methods.  

The Random Forest algorithm has some advantageous properties to address these issues.  First, bagging $-$ where many different models vote on their predictions $-$ has been shown to work well even for problems with considerable classification noise due to incorrect class labeled \citep{Dietterich00}.  Random Forest  should thus be less susceptible than some other classifiers not only to using stars with unknown disposition as negative labels, but also to mix ups between candidate and false positive status in stars labeled as true transits.  Second, the construction of trees with randomized  subsamples of the training sets essentially mimics the cross-validation approach to classification without requiring the removal of objects from the sparse `planet' training set. The  `out-of-bag' (OOB) predictions of the Random Forest algorithm allow improved estimate of node probabilities and error rates in tree construction. Variants of the Random Forest algorithm can also treat class imbalance \citep{Chen04}.

\subsection{Receiver Operating Characteristic (ROC) Curves}
\label{sec:roc}

\begin{figure}[t]
  \label{fig:roc_schematic}
  \begin{center}
    \includegraphics[height=4in]{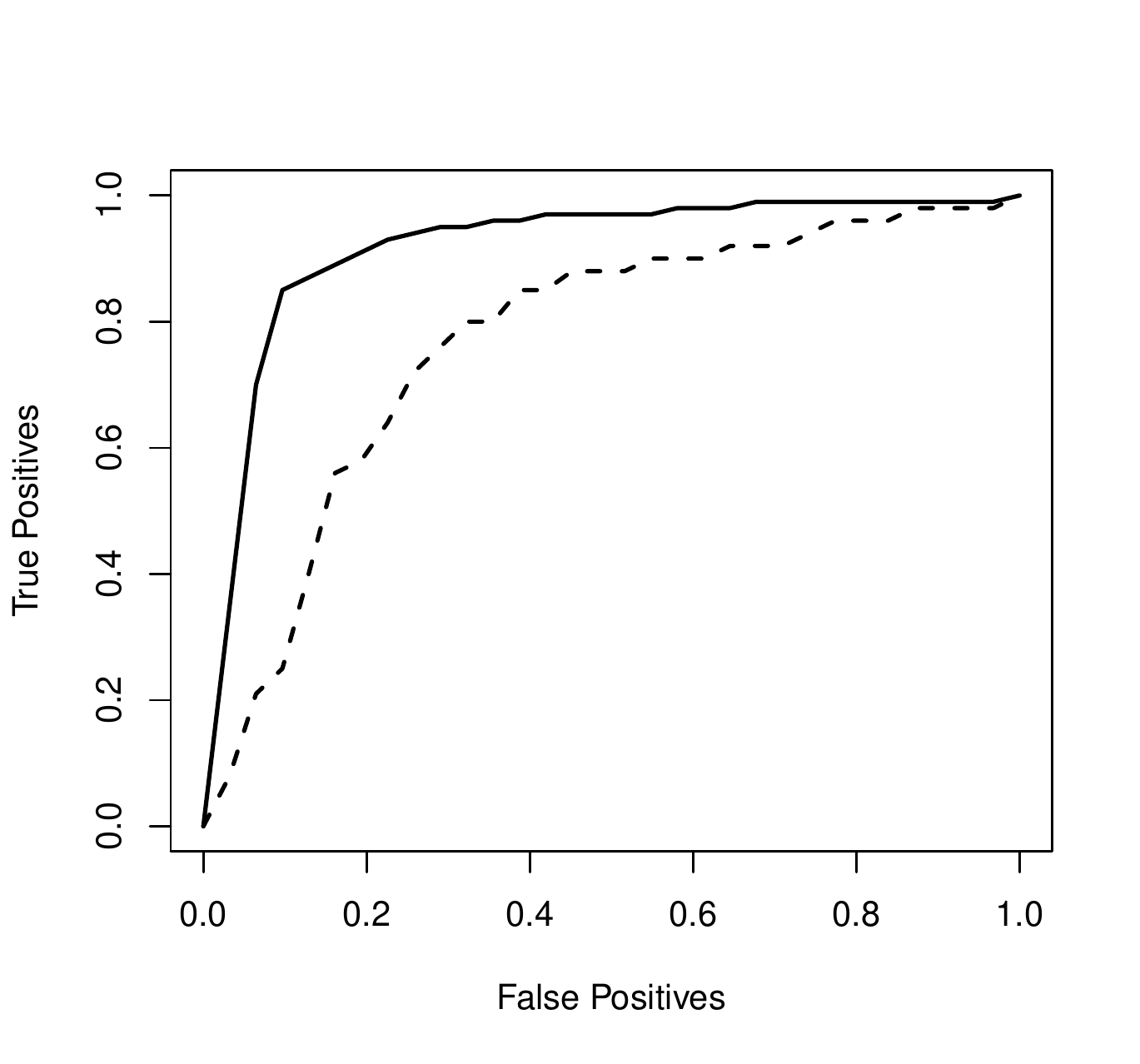}
    \caption{Schematic view of the classification stage for identifying candidate transiting planets.  ROC curves for a single parameter measuring the peak in the TCF periodogram (dashed curve), and for a classifier combining the predictive effects of multiple features of the light curve, periodogram and folded light curve (solid curve). } \label{fig:ROC}
  \end{center}
\end{figure}

Many approaches, whether univariate or multivariate, perform ``soft'' classification through a real-valued output such as periodogram power or Random Forests probability. When seeking to predict a specific category (``hard'' classification), for example to create a candidate exoplanet catalog or list for follow-up telescopic study, a cutoff threshold must be defined. For the case of discrimination between two classes using a single real-valued feature, ROC curves are a useful tool for both evaluating a classifier's performance and selecting a threshold value~\citep{Hanley1982, Fawcett2006, Krzanowski2009}. ROC curves (\S\ref{sec:roc}) allow us to compare the trade-off between the False Positive Rate (FPR) and the True Positive Rate (TPR) as a function of the cutoff value of some classificatory measure, such as the probability emerging from a Random Forest classifier or the SNR of the peak values in a periodogram.  The TPR is the ratio of true positives recovered by the classifier to all events which are actually positive, and similarly, the FPR is the ratio of recovered false positive to all events are in truth negative.  

One important use of ROC curves is to compare the classificatory effectiveness of different classifiers (e.g., using different methods, variables, or training sets).  The scalar measure called  Area Under the Curve (AUC) is often used in this capacity~\citep{Bradley1997}.  The AUC can be  interpreted as the probability that a randomly chosen positive case is ranked higher by the algorithm than a randomly chosen negative case. 

For illustration purposes, Figure~\ref{fig:ROC}  shows an example ROC curve which could be used to compare two different classification procedures from a classifier of transit $vs.$ non-transit light curves. The ideal location is the top-left corner of the plot, where $TPR=1$ and $FPR=0$. Using known candidates, we can calibrate the metric of interest to a desired combination of TPR and FPR.   Several scalar metrics that combine TPR, FPR and related quantities are used including: sensitivity, specificity, precision, false discovery rate, F1 score and Matthews correlation coefficient\footnote{See \url{https://en.wikipedia.org/wiki/Precision_and_recall}.}.  This last quantity is designed to be less sensitive to imbalances in training set sample sizes.  

\section{Discussion}
\label{sec:disc}

The AutoRegressive Planet Search (ARPS) statistical procedure has three main stages: ARIMA-type modeling of the light curve; TCF periodogram to find periodic variations in the model residuals; and Random Forest classification of stars based on features tuned to discriminating a training set of confirmed exoplanetary light curves from non-exoplanetary light curves.  We discuss these procedures in the following three subsections; a fourth subsection discussing limitations and possible improvements to ARPS analysis. 

\subsection{ARIMA Light Curve Modeling} \label{sec:disc_arima}

Autoregressive models are a rich and flexible family of time series models describing stochastic processes.  ARMA models are designed to describe the behavior of stationary autocorrelated  time series.  They have been used by astronomers to model some time series, particular CARMA models of quasar variations \citep{Kelly2014}.  But they are formally valid only for stationary processes with constant mean and variance.  ARIMA models extend these to deal with many forms of nonstationarity. Since stars (e.g., rotationally modulated starspots) and accretion processes can be nonstationary, ARIMA should be preferred to ARMA for most astronomical time series studies.  ARIMA models autocorrelation in the \textit{changes} of the series (for the case of a single differencing operation, the ARMA process is calculated for point-to-point differences in the time series). ARFIMA models allow for \textit{fractional} differencing in order to model long-memory autocorrelation.  The model is multiscale in the sense that the AR and MA components treat small-scale variations while the I and FI components treat long-timescale variations. ARIMA-type  models have shown to be very effective on a variety of applications and fields, and  these successes motivated us to explore their application in astronomy where they have not seen widespread use \citep{Feigelson2018}.

In addition to their success in practical applications, autoregressive models also have a solid theoretical foundation motivating their use. The Wold Decomposition Theorem \citep{Wold1938} guarantees that any stationary, infinite-time, stochastic time series can be decomposed into a linear combination of the random innovations driving the system. There is a deep connection between AR and MA processes and thus ARMA is a parsimonious approximation to the decomposition presented by the Wold Theorem~\citep{Chatfield2004,Shumway2006}. However, the theorem does not promise that a given time series can be accurately modeled by low-dimensional (vs. infinite dimensional) ARMA(p,q) models. This can only be demonstrated by application to specific datasets under study.  

Representing the data in this way can be seen as analogous to a Fourier decomposition, which is much more commonly used in astronomy. As discussed by~\citet{Scargle1981}, while frequency domain techniques are quite powerful when dealing with harmonic variations they are not as effective with random variations.~\citeauthor{Scargle1981} further describes the applicability of time-domain analysis to astronomy and indicate scenarios where it may be superior to frequency domain techniques.~\citet{Koen1993} and \citet{Feigelson2018}  also detail the use of ARMA-type models for astronomical time series.

Even if a light curve looks seemingly white, or has low variability (quantified by IQR), it is still possible for it to have autocorrelated noise. The tools presented in \S\ref{sec:acf} help evaluate whether autoregressive models may be useful for a particular dataset. When a time series is already white, the application of the differencing operator can produce a anti-correlation at a lag of 1 and increase the IQR slightly; this is indicative of `overdifferencing' \citep{Ruppert10}. Overdifferencing can reduce the efficiency of the estimates, but it should not lead to serious inference errors.  The dangers of overdifferencing are fewer than that of insufficient differencing \citep{Plosser77, Harvey81}. The lag=1 structure added during overdifferencing is typically removed in the ARMA modeling that immediately follows. In ARPS, a forced single-order differencing operation is needed to obtain uniform double-spike patterns for planetary transits (\S\ref{sec:tcf}).  The subsequent ARMA modeling will remove most of any autocorrelation induced by overdifferencing. 

Although these models describe stochastic processes, the parameter estimation follows a deterministic regression procedure based on maximum likelihood estimation. The residuals from the fit correspond to the best estimate of the random innovations affecting the system.  The light curve fitted values from the model are always conditioned on the previous known data; that is, they correspond to the one-step in-sample forecast. 

Another strength of ARMA models is that they are not limited by missing data and require no imputation of missing values. The maximum likelihood can be estimated exactly via Kalman filtering even in the presence of missing values~\citep{Gardner1980}.  This gives the flexibility to model irregularly spaced ground-based astronomical light curves where the cadence is not too sparse.  The measurement can be binned into to fixed time grid with `Not Available' entries during daylight or other gaps in the data stream.  This approach for exoplanet transit discovery in light curves with irregular cadences is examined by \citet{Stuhr2019}.  Additionally, ARMA-type models can be very effective for imputation of missing data for stellar light curves with gaps  \citep{Moritz2017}.

In our application to $\simeq$200,000 Kepler light curves \citep{Caceres2019b}, we are impressed by how well ARIMA models perform on a wide variety of stellar variability using very few parameters.  Indeed, the problem can arise that the fits to the stellar variability are too accurate;  some of the planetary transit signal can be incorporated into the model and removed from the residuals that are subject to TCF analysis.  Nonparametric modeling of stellar light curves can encounter the same problem of `overfitting' the stellar variations.  In such cases, the scientific motivation of discovering faint planetary signals may warrant examining residuals of less accurate models, such as an ARIMA rather than an ARFIMA model.  

\subsection{TCF Transit Search}

After stellar variability has been significantly reduced from the time series with autoregressive modeling, we are left with the main task of finding planetary transits in the model residuals (\S\ref{sec:tcf}).  In accord with the analysis of \citet{Kovacs2016}, we treat the search for periodic transits as a stage of analysis distinct from stellar variability reduction, as there are disadvantages of combining these stages.   We adopt the basic approach of Box-fitting Least-Squares  by applying a matched filter to a simplified box-shaped transit.  In the case of Gaussian noise, this is the optimal, maximum likelihood solution.   We introduce the Transit Comb Filter (TCF) algorithm to account to treat the transformation of a box-shape into a double-spike, as shown in Figures~\ref{fig:diff-box}-\ref{fig:fold-ex}.  Both the BLS and TCF  algorithms loop over possible periods, phases, and durations to select the combination that provides the smallest square error with respect to the data. 

Equation~(\ref{eq:mse}) frames the optimal filter as the minimum square error solution, which then is restated as a maximization problem in equation~(\ref{eq:tcf}).  Working with an equally-spaced time series gives a discrete set of values to estimate the transit which when, combined with our specific filter shape, can be used to speed up computation relative to the standard BLS approach. This shown in the simplification of equation~(\ref{eq:mse-final}) to equation~(\ref{eq:tcf}). Furthermore, we do not need to define an arbitrary number of bins in the folded light curve, although folding is necessarily discretized by the observational cadence.

The ultra-short period search performed by~\citet{Sanchis-Ojeda2014} bears some conceptual similarities to the approach underlying our TCF algorithm. Using a simple fast-Fourier transform, they searched for ultra-short period planets using the fact that a signal composed of many, short-period spikes in the time domain has its information condensed into just a few higher-power spikes in frequency-space. This stems from the fact that the Fourier transform of a Shah ($\Shah$) function is also a Shah function with scaling and period inversely proportional to the period of the original.  Their approach focuses on finding the single highest peak in the spectrum, but one can imagine using a filter that corresponds to the exact expected shape just as it can be done in the time domain. The TCF algorithm then has a similar conceptual premise, with the difference that we are explicitly doing the convolution of the Shah function with the double spike in the time domain, and can do so efficiently due to the nature of our filter. This connection helps justify our sensitivity to ultra-short period planets, combined with higher sensitivity to longer periods compared to a simple Fast Fourier Transform at the expense of an increase in computational load.

\subsection{Random Forest Candidate Selection}

When a labeled training set with confirmed planets is available, classification algorithms can be used to improve candidate selection compared to a single scalar value such as periodogram power or SNR. Ideally, sensitivity to weak candidates is increased while simultaneously decreasing the number of false positives. This can be achieved by incorporating additional information of the star and from other stages of the analysis such as stellar size or magnitude,  variability characteristics of the original and residual light curves, properties of the TCF periodogram and folded light curve, and estimated transit parameters. A multivariate decision tree then separates classes taking into account information from these features.  Decision criteria need not be simplistic; for example, different periodogram power thresholds may be effective for different period ranges.  An important operational question is whether outliers might be removed before a characteristic is measured or a criterion is applied. 

In an idealized setting, an algorithm would `learn' features directly from light curves or periodograms; this is the approach of \citet{Pearson2018} \citet{Shallue18} and \citet{Zucker2018} who use convolutional neural networks for exoplanet detection that train directly from the original light curves.  But  we take the more common (and less computationally intensive) approach that requires the scientist to develop a catalog of scalar quantities that assist in classification; `feature engineering' in machine learning parlance.  While it can be challenging to select features that summarize important attributes in a light curve or periodogram, classifiers like Random Forests are effective at ignoring irrelevant or redundant variables in constructing a classificatory structure.  

For the planetary transit detection problem, the choice of training sets is also not obvious.  While one clearly wants {sets of `confirmed planets' and `confirmed non-planets' without classification errors, it is not obvious how to deal with the various types of instrumental and astronomical false positives that confound the sample.  The most difficult contaminating population is blended eclipsing binary systems whose light curves can be very similar to planetary transits.  Known false positives could be incorporated into the `confirmed non-planets' training set, or collected into a third training set.  Furthermore, the transit classification problem is extremely `imbalanced' where the number of non-planetary light curves can exceed the true planetary light curves by factors of dozens, hundreds, or thousands.  As statistical classifiers perform best with balanced training sets, careful choices of sample sizes and stratification are needed in designing the training sets.
 
Our work is similar in some respects to other applications of machine learning methods to transiting exoplanet discovery.  Random Forests have been trained to differentiate between planet candidates (signal consistent with transiting planet), astrophysical false positives (signals like eclipsing binaries or starspots which can resemble planets), and non-transiting phenomena (spurious signals due to the instrument or other source of noise) by \citet{McCauliff2015} and \citet{Catanzarite2015}. Other techniques, such as \textit{k}-nearest neighbors, have also been used~\citep{Thompson2015}.
 
Random Forests have the advantage of being metric-free so that `distances' in multivariate space are not calculated.  It is thus not influenced by varying scales and ranges of different features unlike many other types of learning algorithms. This is related to `affine invariance', an effective approach to a variety of pattern recognition and simulation problems~\citep{Foreman-Mackey2013}.  Data on many different scales and wide range of values are common in astronomy, and this makes Random Forests particularly appealing for these applications. Care must be noted that certain biases do exist towards features with a larger number of unique values, such as between categorical predictors containing different number of classes or relative to continuous variables~\citep{Quinlan1987}.

An additional challenge for automatic classification is the proper validation of performance. Standard practice is to use some form of test set, distinct from the set used to train the algorithm, to validate the results. This is often in the form of a holdout set (e.g., 20\% of the data is withheld from analysis) or cross-validation (where the data is split into multiple subsets and retrained, each time leaving a subset out of training to be used only for evaluation). Random Forests have the useful property that each tree is built on a bootstrapped sample (random draws with replacement) of the data, leaving different instances out of the training set in each tree of the forest. This method, particularly the `out-of-bag' error estimate within Random Forests, provides a natural set to evaluate the model's performance \citep{Breiman2001}.   However, biases are possible and cross-validation may provide a reasonable check on classification validation.  Expert opinions differ on this matter.  

Since ROC curves, the AUC and related scalar measures of classification success are estimated based on the relative scores between positive and negative instances, discrimination does not depend on the value being a properly calibrated probability~\citep{Fawcett2006}. This avoids the concern of whether the probabilities estimated by, for example, a Random Forest are ``true'' probabilities, and also enables use of uncalibrated features (like periodogram power and SNR with non-Gaussian distributions) for classification. The distinction also emphasizes why we do not default to a fixed cut-off like 0.5 or 0.9 as would be expected in the case of a true probability estimate. These tools are particularly useful when classifying highly imbalanced data. Both TPR and FPR are \textit{rates}, each respectively scaled to the number of positive and negative cases in the sample, irrespective of their relative occurrences, and thus insensitive to any imbalance in the validation data.  

However, it is not obvious what criterion should be used to choose the `best' threshold from a ROC curve when training sets are highly imbalanced.  Consider the inappropriate use of classification ``accuracy'', defined as the ratio where the sums of true positives and true negatives is in the numerator, and the total population is in the denominator.  When one class greatly outnumbers another, a classifier could learn to simply predict the dominant class for any future instance. While this would give a very high accuracy, it would mean no observation would ever be classified as belonging to the class with fewer members even though those may be the most interesting ones, as in the case of discovering new planets.

Multiple methods exist to adapt or tune a classifier's performance on imbalanced data. Graphical tools, such as a precision-recall plot, can be helpful.  Matthew's correlation coefficient, an arithmetic combination of True Positive and False Positive counts related to the $\chi^2$ statistic for a $2 \times 2$ contingency table, is particularly recommended when the populations of the classes are highly imbalanced \citep{Boughorbel2017}. Another approach is to define a loss function that avoids mistakes such as the accuracy example presented above. The cross-entropy loss (also known as the logarithmic loss used in logistic regression) highly penalizes incorrect classification made with a high degree of certainty and thus can help veer away from incorrectly classifying the smaller represented class. 

\subsection{Limitations \& Possible Improvements to KARPS Methodology}

The methodology presented in this work has a number of limitations.  Some are intrinsic to the methodology, and others can be ameliorated in various ways.  
\begin{description}

\item[Model misspecification] ARIMA models are low-dimensional, linear, parametric models. As such, they may not always correctly describe complex stellar behaviors. Residual analysis (including tests for normality, autocorrelation and stationarity) can help assess whether model misspecification is issue. In \citet{Caceres2019b} where the full dataset of $\sim$200,000 Kepler stars is analyzed using ARPS methods, we find a significant minority of systems are not well-fit with ARIMA or ARFIMA models.  It is possible that some types of variability, such as those caused by sudden high amplitude shocks, will not have simple autoregressive properties.  In such cases, the ARPS approach to planet search may not be very effective.

\item[Equally-spaced data] The mathematics of autoregessive modeling discussed in this paper has been developed for evenly-spaced data~\citep{Hamilton1994}, but most astronomical time series are irregularly spaced. Two approaches exist to extend ARMA-type methodology for irregularly sampled observations:  resample the data into an equally-spaced grid since ARIMA models can treat missing data, or use extensions of the methodology to handle continuous processes~\citep{Jones1985, Feigelson2018}.  The first approach is examined for the planetary transit problem by \citet{Stuhr2019}.  The second approach based on continuous-time ARMA (CARMA) models have been used to characterize quasar variability from ground-based photometric surveys \citep{Kelly2014}.  A new CARFIMA software implementation is now available that provides a richer family of variability models than CARMA \citep{Tak2017}.  The relative merits of these two approaches have not yet been evaluated.

\item[Normal error distribution] These models assume the errors are homoscedastic and normally distributed. Extensions such as GARCH (Generalized Autoregressive Conditional Heteroscedasticity) exist to treat autoregressive heteroscedastic scatter \citep{Hyndman2014, Enders2014, Greene2017}.  In our examination of $\simeq$200,000 Kepler stars, we found some light curves that exhibited volatility that benefitted from GARCH modeling, but this was not common. 

\item[No weighting by measurement error] All scatter is incorporated into a single model component, such as $\epsilon$ in equation~(\ref{eq:arima}), without distinguishing between measurement error and intrinsic stellar variability. Furthermore, as a stochastic process, it is implicit in the model that the random effects can influence its evolution.  A hierarchical state space formalism for parametric time series modeling, including both stochastic and deterministic processes,  could be adopted in order to explicitly separate measurement and intrinsic errors~\citep{Durbin2012, Casals2016}.

\item[Nonstationarity and de-trending]
ARMA models require that the data be stationary (which implies reverting to a constant mean value over time). In practice, we use the differencing operator in ARIMA to achieve approximate stationarity. However, the formal mathematical use for differencing is to treat \textit{stochastic} trends (such as a random walk), and deterministic trends should be modeled separately. When the star varies in a more complex manner than simple differencing can handle, one can preprocess the light curve by detrending using a semi-parametric density estimator such splines, LOESS, or a Gaussian Processes regression fit.   

\item[Periodicities] 
A particular form of nonstationarity, such as seen in Cepheid variables and eclipsing binaries, are strictly periodic variations. ARIMA does not model these behaviors well, as it is designed for stochastic behaviors.  Frequency domain methods, such as Lomb-Scargle periodograms or Stellingwerf's phase dispersion minimization,  can help identify strictly periodic phenomena and allow it to be removed (pre-whitening) prior to ARIMA modeling. A periodic component with unknown or uncertain amplitude, but with known period and phase, can be treated as an exogenous variable in an extended ARIMA model, as with our evaluation of TCF periodogram peak significance in section~\ref{sec:arbox}.  The ARMAX formulation allows simultaneous inference of periodic or trend parametric model parameters with stochastic autoregressive model parameters.

\item[Filter shape] A box-shaped transit and the corresponding double-spike after ARIMA modeling, are simplified versions of a transit. A straightforward extension to the TCF algorithm can use weighted spikes when a transit's ingress and egress is split among multiple cadences. A more advanced and astrophysically motivated transit model could be implemented to further improve sensitivity, such as the \citet{Mandel2002} transit model with stellar limb darkening. Simultaneous inference of ARIMA and an astrophysical model may require a hierarchical state space model \citep{Durbin2012}.  Finally, in ARFIMA modeling, due to the fractional differencing, the transit shape is distorted in a more complex way than shown in Figure~\ref{fig:tcf-sketch}, so the TCF designed here for periodic double-spike detection is no longer an optimal matched filter\footnote{We are grateful to Prof. Soumendra Lahiri (Statistics, NCSU) for raising this last point.}.   
\end{description}

In addition to the mathematical restrictions above, several astronomically motivated concerns arise in ARPS applications. High-resolution imaging and spectroscopic followup are often needed to confirm that orbiting planets are truly present, even when a light curve satisfies the classification requirements demonstrating close resemblence to other exoplanetary systems. 
\begin{description}

\item[Instrumental effects]  Often the astronomer has access to ancillary data regarding telescope or detector performance, or regarding atmospheric conditions.  These can be used to pre-process the light curves period to ARPS analysis (footnote~\ref{footnote2}).  For example, for the Kepler dataset \citet{Caceres2019b} use star fluxes calculated after a complicated Pre-Search Data Conditioning pipeline has been applied.  

\item[Stellar Characteristics] We do not explicitly account for astronomical aspects of each observed source, such as: distance; stellar type (e.g., dwarf, giant); stellar properties (temperature, size); and any related constraints on transit properties.

\item[Blending] Spatial blending of background or foreground eclipsing binaries can contaminate measurements and lead to astronomical false positives~\citep{Torres2004}, which are not examined in this work. However tests do exists for more detailed investigation of promising candidates~\citep{Torres2011}.

\item[Transit signal reduction] As mentioned in \S\ref{sec:disc_arima}, the concern exists that the differencing operation and autoregressive modeling might overfit the stellar variability and eat away the planetary transit signal.  This limitation applies to any method trying to remove the underlying noise, such as wavelet analysis or Gaussian Processes regression. In our explorations, signal loss is typically offset by the noise reduction, leaving a net positive gain in sensitivity.  In the Kepler sample, we find that the more flexible ARFIMA model is so successful at  reducing stellar variations compared to the less flexible ARIMA mode that is also reduces the transit signal in the TCF periodogram \citep{Caceres2019b}.  

\item[Multiple planet system]  We have not developed the ARPS procedure to treat two or more transiting planets; the secondary planets are ignored here.  However, a procedure of prewhitening the light curve of the primary planet signal, and then recalculating the TCF periodogram to identify additional periodic signals, can readily be developed.  
This can be iterated until the TCF peaks are sufficiently weak that the Random Forest classifier considered it to be a non-planet.  

\end{description}

\section{Conclusions}
\label{sec:concl}

We present here a statistical procedure for identifying planetary transits in stellar light curves, nicknamed Autoregressive Planet Search (ARPS).  It is founded on parametric regression models such as ARIMA that flexibly model time series with complicated autocorrelated and trend behaviors at low dimensions. Best-fit ARIMA models are calculated by maximum likelihood estimation with no free parameters, followed by regression diagnostics to evaluate the success of the model.  The goal is to reduce unwanted aperiodic stellar, instrumental or atmospheric variations to better reveal faint periodic planetary transit signals.  

Planetary transits are then sought in the model residuals using periodogram based on a novel matched filter algorithm we call the Transit Comb Filter.  The calculation, similar to the Box-Least Squares algorithm, is computationally efficient.  The light curve of the model residuals folded at the period of the TCF periodogram peak is characterized, and the ARIMA model is calculated again with this periodicity as an exogenous variable.  

At this stage, the scientist can examine the results---original light curve, TCF periodogram, folded light curve, and derived quantities---to subjectively identify likely exoplanetary transit signals.  The difficulty is rejection of various types of false positives, particularly blended eclipsing binaries with periodic behaviors that can mimic planetary transits.  If, however, some planets have already been confidently identified from the dataset under study, they can be used as a training set for a multivariate classifier.  We use decision trees and Random Forests based on a collection of features drawn from the original light curve, TCF periodogram, and folded light curve.  With visualizations and scalar criteria based on ROC curves, we find that this multivariate classification procedure is considerably more effective at detecting faint planetary transits and reducing false positives than a threshold based on a univariate TCF periodogram peak measure \citep{Caceres2019b}.  

Our objective is to introduce an approach to transit detection that is complementary to those in common use.   ARIMA-type analysis of non-planetary stellar variations are modeled using low-dimensional {\it parametric} models, not non-parametric or semi-parametric approaches like wavelet decomposition or Gaussian Processes regression.  We do not believe that any single method will out-perform the others in all cases.   Rather, different methodologies can capture different behaviors of complex light curves. Autoregressive modeling has strengths such as unique maximum likelihood solutions, AIC-based approach to model complexity, and parameter confidence intervals.  The sequence of ARIMA-type modeling with TCF periodograms and Random Forest classification is a particularly effective combination of methods for the specific goal of transiting planet detection.

The basic ARPS procedure is more clearly defined than some other common approaches. Once an ARMA-type family is chosen (e.g. ARIMA, ARFIMA, ARIMAX), a unique maximum likelihood model is obtain for a given order, ARIMA(p,d,q), and an optimal choice of order is obtained using the Akaike Information Criterion for model selection. There are no free parameters to be chosen in the autoregressive modeling stage of the ARPS analysis procedure, and no choice of methods such as Gaussian Processes regression or wavelets.  Construction of the TCF periodogram is also fixed except for computational choices relating to the number of phases and teeth width considered.  The classification stage is more open-ended with scientific judgment needed for feature selection, thresholds based on ROC curve, and subjective rejection of  remaining false alarms and false positives that satisfy the classification criteria.

This paper lays the foundation of the ARPS method for several studies in progress:
\begin{enumerate}

\item We are applying ARPS to the space-based Kepler mission dataset with 4 years photometry of $\sim$200,000 stars \citep{Caceres2019b}.  The principal goal is to identify new candidate exoplanets that are similar to the DR~25 Golden sample of \citet{Twicken2016} which serves as a training set. The result is the identification of several dozen new candidate transit systems, particularly with very short orbital periods.

\item We are investigating the ARPS transit detection procedure for irregularly spaced time series produced by ground-based photometric surveys where the noise characteristics are dominated by atmospheric and instrumental effects, rather than by stellar variability \citep{Stuhr2019}.  Although ARIMA modeling is designed for evenly spaced time series, we find reasonable sensitivity to planets providing the observing cadence is sufficiently dense.  Application to the HAT-South photometric dataset is underway.  

\item The potential role of ARIMA-type modeling for addressing issues in a broader range of time domain astronomy astronomy is reviewed by \citet{Feigelson2018}.  Potential applications include research on variable stars and accretion-dominated systems such as cataclysmic variables and quasars, in addition to exoplanet detection.  
\end{enumerate}

\acknowledgements  
We appreciate valuable discussions with  Suzanne Aigrain, Eric Ford, Ron Gilliland, Jon Jenkins, Angie Wolfgang, and Jason Wright as these methods were developed. Two anonymous referees provided helpful commentaries.   EDF and GJB are affiliated with Penn State's Center for Astrostatistics. M.C. thanks the support from Centro de Astrofisica de Valparaiso and Centro Interdiciplinario de Estudios Atmosfericos y Astroestadistica.  This work is supported by NSF grant AST-1614690 and NASA grant 80NSSC17K0122 at Penn State.

\end{document}